\documentclass[english,aps,manuscript,nofootinbib]{revtex4}
\usepackage[T1]{fontenc}
\usepackage[utf8]{luainputenc}
\usepackage{float}
\usepackage{amssymb}
\usepackage{graphicx}
\usepackage{esint}

\makeatletter

\providecommand{\tabularnewline}{\\}

\@ifundefined{textcolor}{}
{%
 \definecolor{BLACK}{gray}{0}
 \definecolor{WHITE}{gray}{1}
 \definecolor{RED}{rgb}{1,0,0}
 \definecolor{GREEN}{rgb}{0,1,0}
 \definecolor{BLUE}{rgb}{0,0,1}
 \definecolor{CYAN}{cmyk}{1,0,0,0}
 \definecolor{MAGENTA}{cmyk}{0,1,0,0}
 \definecolor{YELLOW}{cmyk}{0,0,1,0}
}

\makeatother

\usepackage{babel}
\begin{document}

\title{Deep Saturated Free Electron Laser Oscillators and Frozen Spikes}

\author{P. L. Ottaviani, S. Pagnutti}

\email{simonetta.pagnutti@enea.it}

\affiliation{ENEA - Centro Ricerche Bologna, via Martiri di Monte Sole, 4 , IT
40129 Bologna, Italy}

\author{G. Dattoli, E. Sabia}

\email{giuseppe.dattoli@enea.it}
\email{elio.sabia@enea.it}

\affiliation{ENEA - Centro Ricerche Frascati, via E. Fermi, 45, IT 00044 Frascati
(Roma), Italy}

\author{V. Petrillo}

\email{vittoria.petrillo@mi.infn.it}

\affiliation{Universita' degli Studi di Milano, via Celoria 16, IT 20133 Milano,
Italy}

\affiliation{INFN - Mi , via Celoria 16, IT 20133 Milano, Italy}

\author{P. Van Der Slot}

\email{p.j.m.vanderslot@utwente.nl}

\affiliation{University of Twente, P.O.Box 217, 7500 AE Enschede (The Netherlands)}

\author{S. Biedron and S. Milton}

\email{biedron@anl.gov}
\email{milton@engr.colostate.edu}

\affiliation{Department of Electrical and Computer Engineering Colorado State
University (USA)}
\begin{abstract}
We analyze the behavior of Free Electron Laser (FEL) oscillators operating
in the deep saturated regime and point out the formation of sub-peaks
of the optical pulse. They are very stable configurations, having
a width corresponding to a coherence length. We speculate on the physical
mechanisms underlying their growth and attempt an identification with
FEL mode locked structures associated with Super Modes. Their impact
on the intra-cavity nonlinear harmonic generation is also discussed
along with the possibility of exploiting them as cavity out-coupler.
\end{abstract}
\maketitle

\section{Introduction}

In a Free Electron Laser (FEL) oscillator the optical pulses go back
and forth inside the cavity and, at the undulator entrance, overlap
with fresh electron bunches, reinforcing the relevant intensity, until
gain equals the cavity losses\cite{Colson}. The pivoting parameters
of the game are defined below

\begin{eqnarray}
 &  & \Delta=N\lambda_{s}\equiv\:Slippage\:Length\nonumber \\
 &  & \sigma_{z}\equiv\:Bunch\:Length\label{eq:eq1}
\end{eqnarray}

\begin{flushleft}
with $N,\:\lambda_{s}$ being the number of periods of the undulator
and the resonant wavelength, respectively. The physical origin of
$\Delta$ is associated with the different velocities of electron
and radiation packets, determining an advance of the radiation, after
an undulator passage, by just a slippage length. For an appropriate
quantification of the effects induced by such a longitudinal mismatch
the following parameter has been introduced \cite{Dattoli-Renieri} 
\par\end{flushleft}

\begin{equation}
\mu_{c}=\frac{\Delta}{\sigma_{z}}\label{eq:eq.2}
\end{equation}

\begin{flushleft}
which has a manifold role. It provides indeed a measure of the number
of longitudinal modes, locked during the interaction and the gain
reduction itself induced by the spoiled longitudinal overlapping.
To understand the reason why the parameter $\mu_{c}$ is associated
with mode locking, we just remind that the longitudinal mode gain
of a FEL driven by a short pulse electron beam is linked to the Fourier
Transform of the pulse itself \cite{Dattoli-Renieri} and discuss
the relevant details in appendix A. 
\par\end{flushleft}

\begin{flushleft}
Slippage and mode coupling cannot be considered disentangled and should
be viewed within the context of the short pulse oscillator dynamics,
also characterized by the lethargy mechanism, due to the slowing down
of the optical packet velocity, while interacting with the electrons
inside the optical cavity. As a consequence either the gain and the
cavity equilibrium power depends on the cavity detuning parameter 
\par\end{flushleft}

\begin{equation}
\theta=\frac{4\,\delta L}{g_{0}\Delta}\label{eq:eq.3}
\end{equation}

\begin{flushleft}
where $\delta L$ is the cavity mismatch from the ideal length \footnote{By ideal cavity length we mean that corresponding to a round trip
allowing the overlapping between two successive bunches of electron
and photons.}, introduced to compensate the lethargy effect and $g_{0}$ is the
small signal gain coefficient.The intra-cavity intensity round trip
evolution can be reproduced by fairly simple formulae \cite{Datt-Ottavio},
reported below. 
\par\end{flushleft}

\begin{flushleft}
\begin{eqnarray}
 &  & I_{r}\left(\theta,\mu_{c}\right)=I_{0}\frac{\left\{ \left(1-\eta\right)\left[G\left(\theta,\mu_{c}\right)+1\right]\right\} ^{r}}{1+\frac{I_{0}}{I_{e}}\left(\theta,\mu_{c}\right)\left\{ \left[\left(1-\eta\right)\left(G\left(\theta,\mu_{c}\right)+1\right)\right]^{r}-1\right\} },\nonumber \\
 &  & I_{e}\left(\theta,\mu_{c}\right)\approxeq\left(1+\sqrt{2}\right)\left\{ \sqrt{\frac{\theta^{\star}}{\theta}}\:exp\left[\frac{1}{2}\left(1-\frac{\eta}{\left(1-\eta\right)G^{\star}}\,\frac{\theta^{\star}}{\theta}\right)\right]-1\right\} I_{s},\label{eq:eq.4}\\
 &  & G\left(\theta,\mu_{c}\right)=G_{M}\frac{\theta}{\theta_{s}}\left[1-ln\left(\frac{\theta}{\theta_{s}}\gamma_{c}\right)\right],\;G_{M}\approxeq0.85g_{0},\;\theta_{s}\approxeq0.456,\:\gamma_{c}=1+\frac{\mu_{c}}{3},\nonumber \\
 &  & 0\leqq\theta\leqq e\frac{\theta_{s}}{\gamma_{c}},\;G^{\star}=\frac{G_{M}}{\gamma_{c}},\;\theta^{\star}=\frac{\theta_{s}}{\gamma_{c}},\;e\equiv Neper\;number\nonumber 
\end{eqnarray}

\par\end{flushleft}

where $r$ denotes the round trip number,$I_{0}$ the initial seed,
$G$ the small signal gain, $I_{s}$ denotes the FEL saturation intensity,
$I_{e}$ the equilibrium intra-cavity intensity and $\eta$ the cavity
losses. The gain dilution due to the slippage and to the finite bunch
length is provided by $\gamma_{c}$. It should be noted that eq. (\ref{eq:eq.4})
yields the intensity intracavity evolution under the assumption of
low gain, it can therefore be considered valid only for values of
the small signal coefficient not exceeding 0.3. In this approximation
the gain function $G(\theta,\mu_{c})$ is vanishing at zero cavity
detuning, the relevant meaning and limit of validity will be discussed
in the following. For a more accurate parametrization, including high
gain corrections, the reader is addressed to refs. \cite{Datt-Ottavio}
and for further comments to ref. \cite{Nishimori}.

The previous formulae provide a fairly reasonable description of the
FEL oscillator dynamics in terms of laser intensity evolution, but
do not provide any information on the pulse growth and shaping, occurring
during the evolution and on the relevant dependence on the various
parameters.

\begin{figure}[H]
\begin{centering}
$\qquad\qquad\qquad$\includegraphics[bb=0bp 0bp 947bp 792bp,scale=0.4]{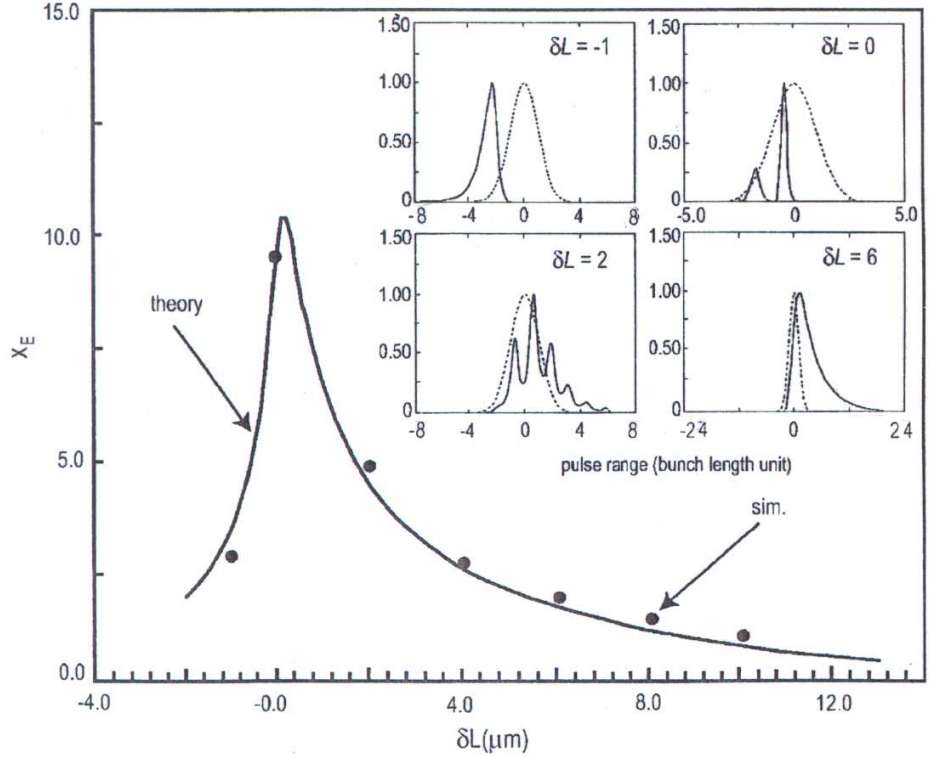}
\par\end{centering}

\protect\caption{\label{Fig1}Oscillator FEL intra-cavity equilibrium power: analytical
(continuous line), numerical (dots) and pulse shape (continuous line)
for different cavity detunings (electron pulse dash line). ($g=2.0,\;\mu_{c}=1,\;\eta=0.06$
. In this case the large gain value allows the growth of the laser
field even for negative values of $\theta$)}
\end{figure}

In Fig.\ref{Fig1} a largely well-known feature is reported regarding
the dependence of the optical pulse shape on the cavity mismatch itself,
the occurrence of different peaks has also been recognized as a manifestation
of the locking of different Super Mode structures \cite{Dattoli-Renieri}.
It should be noted that the concept of Super Mode has been used improperly.
According to the original formulation they are low gain small signal
self reproducing collection of longitudinal modes. They are characterized
by vanishing gain at zero cavity detuning. The fact that they have
no gain at zero cavity desynchronism does not mean that FEL oscillators
cannot lase in this configuration, it only means that for $\theta=0$
there are not stationary modes of S. M. type allowed by the FEL dynamics.
In any case these modes provide a suitable basis allowing the FEL
pulse expansion and eq. (\ref{eq:eq.4}) provides a reasonable approximation
to describe the intensity FEL oscillator evolution in the pulsed regime.
Subsequent elaboration (either numerical and analytical) \cite{Datt-Ottavio,Nguyen}
have clarified the role of the high gain contributions, as we will
further discuss in the following. 

The FEL oscillator pulse dynamics becomes more intrigued with the
system undergoing deeper and deeper saturated regime \cite{Bonifacio}.
One of the most interesting phenomenon occurring at this level is
the emergence of a comb structure inside the optical pulse itself,
consisting of a series of sub-pulses with a length comparable to a
coherence length. 

In this paper we speculate on the physical mechanisms underlying the
emergence of these spikes and discuss the feedback on the intra-cavity
harmonic generation. In the following we will exploit the code PROMETEO
to analyze the evolution of the FEL oscillator dynamics from small
signal to deep saturated regime. The results of the code, which does
not contain three dimensional effects, will be compared with those
from GENESIS, which brings further and deeper insight by allowing
the possibility of understanding the interplay between transverse
and longitudinal modes.

\section{Deep Saturation and optical pulse oscillator dynamics}

\begin{figure}[H]
\includegraphics[angle=360,scale=0.5]{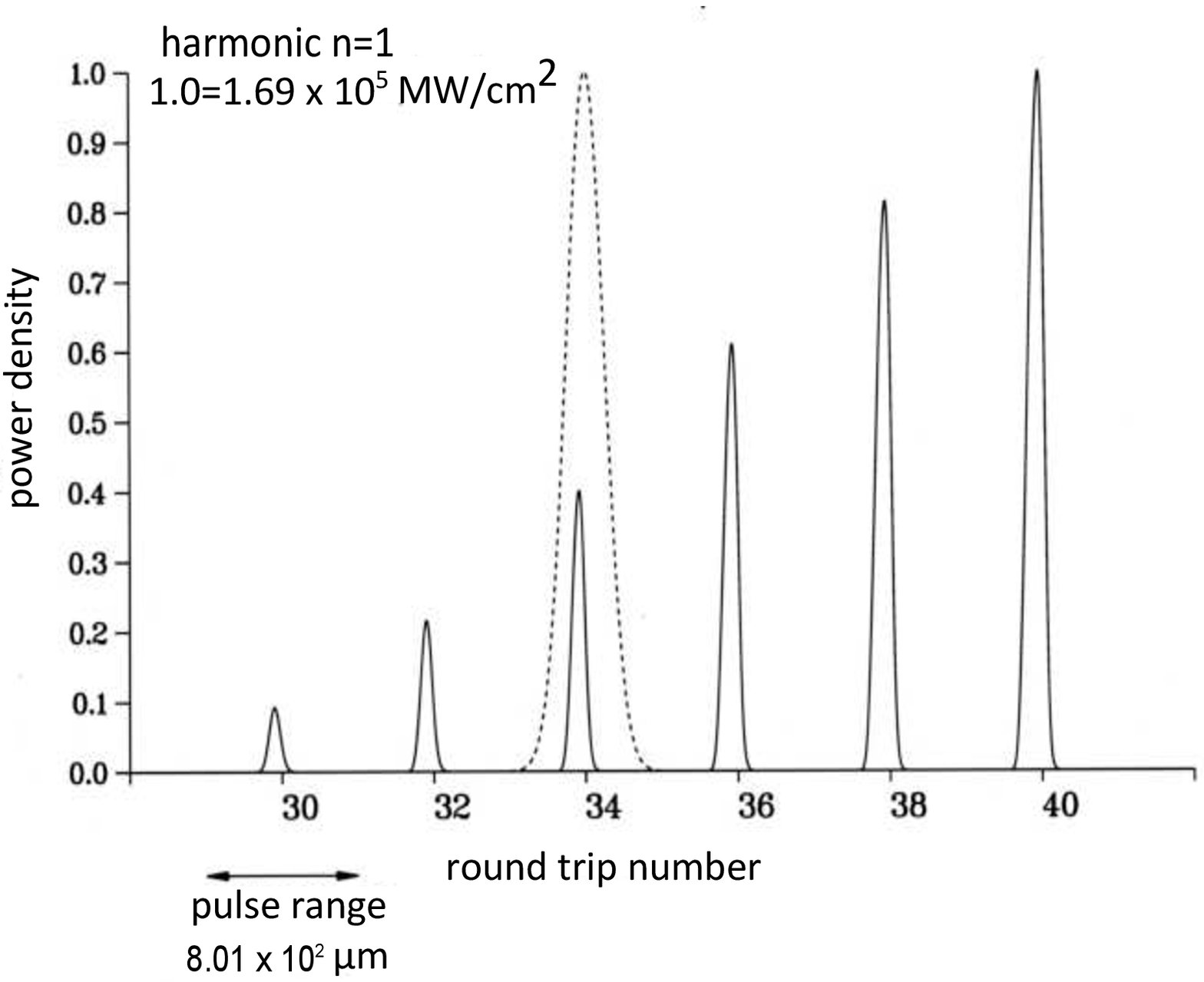}\includegraphics[angle=360,scale=0.5]{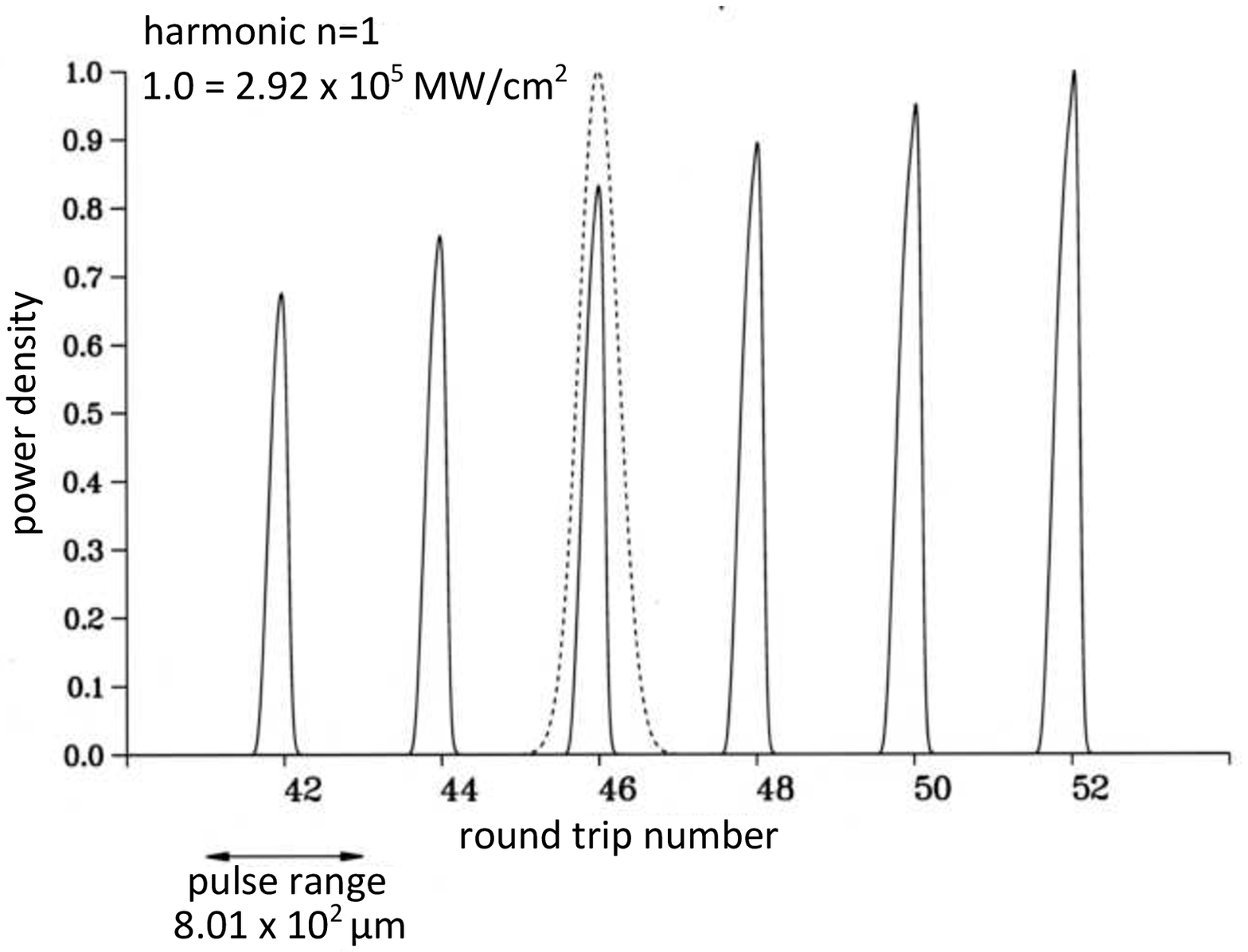}

\includegraphics[angle=360,scale=0.5]{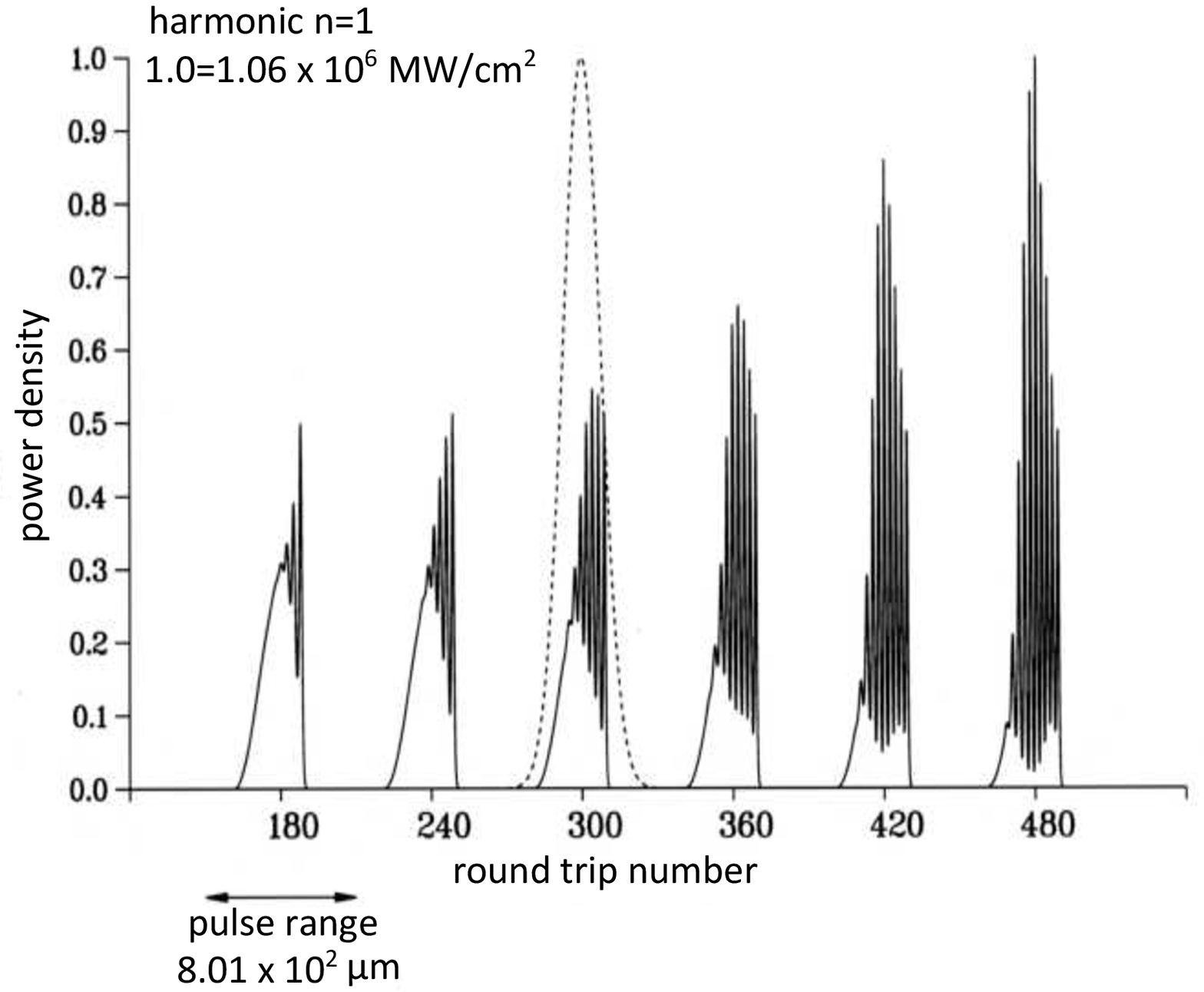}

\protect\caption{\label{Fig.2}Evolution of the optical packet distribution at different
round trip; simulation parameters are given below $E=155.3\:Mev,\;J=2.68\cdot10^{8}\frac{A}{m^{2}},\;\rho=2.33\cdot10^{-3},\;\sigma_{\varepsilon}=10^{-4},\;\sigma_{z}=100\:\mu m,\;g_{0}=1,\;\eta=0.06,\;number\:of\:periods\:N=77,\:\lambda_{s}=500.33\:nm,\:\Delta=38.5\mu m,\;l_{c}=9.78\:\mu m$}
\end{figure}

In Fig.\ref{Fig.2} we report a series of snap-shots describing the
growth of an optical pulse at different round trips, inside the optical
cavity set at zero detuning. The relevant dynamics is, at the beginning,
fairly transparent; the pulses grow in the form of a homogeneous and
almost Gaussian distribution, which over the round trips, tend to
spread over all the electron bunch length. One of the consequence
of the lethargy mechanism is an asymmetry, getting more pronounced
with increasing saturation and characterized by a sharp front edge.

However in the small signal regime and at moderated saturation levels,
the laser-pulse shape is not far from a Gaussian with a bunch length
given by
\begin{equation}
\sigma_{b}\approxeq\frac{1}{2}\sqrt{\Delta\sigma_{z}}\label{eq:eq.5}
\end{equation}

\begin{figure}[H]
$\qquad\qquad\qquad\qquad\qquad$\includegraphics[angle=360,scale=0.55]{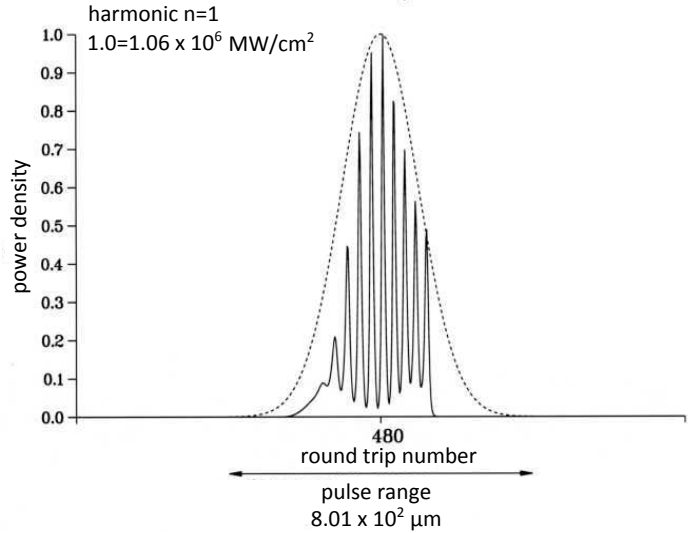}

\protect\caption{\label{Fig.3}Optical packet “equilibrium” configuration; simulation
parameters as in Fig.\ref{Fig.2}}
\end{figure}

While the saturation increases, the bunch acquires a modulation, starting
from the front part and covering progressively the entire electron
bunch, as shown in Fig.\ref{Fig.3}, where we have picked out the
optical pulse distribution at the round trip number 480. For the chosen
parameters of the simulation (cavity length and macro-pulse e-beam
structure) this number of round trip corresponds to tens of microseconds
of time duration. The observation of such a comb structure requires
therefore quite a long operation time.

As already remarked, the slippage length is a measure of the maximum
amount of the optical packet slippage over the electrons, due to the
different velocities. The introduction of such a quantity is however
based on purely kinematical argument, the motion of the optical pulse
inside the electrons cannot be indeed merely viewed as a slippage.
The electrons move indeed in a dispersive medium. The interaction
itself should be characterized by an index of refraction, which in
turn provides the already quoted lethargy mechanism. The FEL refractive
index is a quantity of paramount importance, which is easily specified
in terms of analytical formulae in the small signal regime. When saturation
occurs, the gain (which is essentially linked to the real part of
the refractive index) is decreased by the laser intra-cavity power
itself and the same holds for the imaginary part. The concomitance
of these mechanisms allows the formation of substructures growing
around the coherence length. We have analyzed the peak formation using
different values of the parameters entering the definition of the
FEL evolution process. We have therefore examined cases with different
gain coefficients and bunch length and we have found that the number
of spikes formed in deep saturation does not depend on the number
of undulator periods and seems to be associated with the coherence
length of the process itself, we have therefore made the following
ansatze (the sub-index S stays for \textquotedbl{}Spike\textquotedbl{})
\begin{equation}
n_{S}\approxeq\frac{\sigma_{z}}{\sigma_{S}},\;\;\;\sigma_{S}\approxeq l_{c}=\frac{\lambda_{s}}{4\pi\rho\sqrt{3}}\label{eq:ec.6}
\end{equation}

where $l_{c}$ is the coherence length \cite{Bonifacio} ($\lambda_{s}$
and $\rho$ are the wavelength and Pierce parameter, respectively)
(for further comments see Appendix). 

In the description of the spike formation phenomenology, we have deliberately
used a language and a notation more appropriate to SASE than to oscillator
FEL devices, the peaks we are describing resemble the spikes of the
high gain regime; they become locked in phase by the operating conditions
in the optical cavity. Such an identification is not new, either from
the theoretical and experimental point view \cite{Hajima} and we
will further comment on this point in the concluding section of the
paper. 

In Tab.\ref{tab:Tab1} we have reported the power density of the peaks
of the pulse shown in Fig. \ref{Fig.3} (numerated from the left to
the right). 

\begin{figure}[t]
\includegraphics[bb=3mm 0bp 544bp 489bp,scale=0.6]{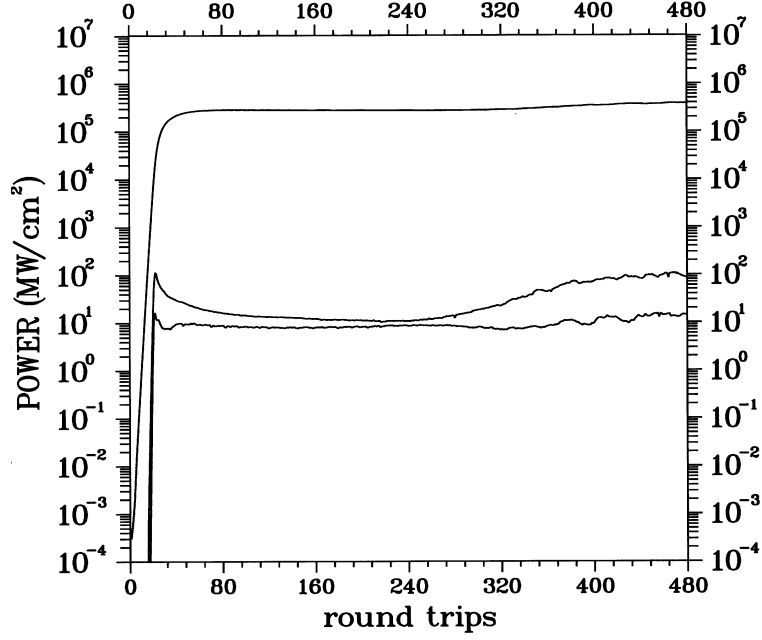}

\protect\caption{\label{Fig4}Fundamental and higher order harmonic (3-rd and 5-th)
power vs. round trip number (the 1-st harmonic is stored, the other
are just radiated coherently at each round trip). Simulation parameters:
(The number of $n_{S}$ calculated with eq. (\ref{eq:ec.6}) for this
specific case is 9) $E=155.3\:Mev,\;J=4.01\cdot10^{8}\frac{A}{m^{2}},\;\rho=2.67\cdot10^{-3},\;\sigma_{\varepsilon}=10^{-4}\:(relative\,energy\:spread),\;\sigma_{z}=100\,\mu m,\;\eta=0.06,\:Number\:of\:undulator\:periods\:N=77,\;\lambda_{s}=500.24\,nm,\;l_{c}=8.66\,\mu m$}
\end{figure}

Before going further let us remind that, along with the fundamental,
higher order harmonics are generated inside the optical cavity. In
Fig.\ref{Fig4} we have reported the power of the first harmonic stored
in the cavity and the power radiated (not stored) at higher harmonics
during each round trip for the case $g_{0}=1.5$.

As it is well known, the non-linear harmonic generation is a by-product
of the FEL process itself and is due to the bunching mechanism induced
by the fundamental harmonics. The maximum of the harmonic emission
occurs before the first harmonic reaches the stationary condition.
In the successive round trips the harmonic power drops down owing
to the large energy spread induced by the FEL interaction itself. 

\begin{table}[H]
$\qquad\qquad\qquad\qquad\qquad\qquad\qquad$%
\begin{tabular}{|c|c|}
\hline 
Peak number & Power density$\left(\frac{MW}{cm^{2}}\right)$\tabularnewline
\hline 
\hline 
1 & $9.32\times10^{4}$\tabularnewline
\hline 
2 & $2.21\times10^{5}$\tabularnewline
\hline 
3 & $4.71\times10^{5}$\tabularnewline
\hline 
4 & $7.85\times10^{5}$\tabularnewline
\hline 
5 & $1.01\times10^{6}$\tabularnewline
\hline 
6 & $1.06\times10^{6}$\tabularnewline
\hline 
7 & $8.71\times10^{5}$\tabularnewline
\hline 
8 & $7.37\times10^{5}$\tabularnewline
\hline 
9 & $5.95\times10^{5}$\tabularnewline
\hline 
10 & $5.17\times10^{5}$\tabularnewline
\hline 
\end{tabular}

\protect\caption{\label{tab:Tab1}Peak number and relevant power at the 480th round
trip.}
\end{table}

\begin{figure}[H]
$\qquad\qquad\qquad\qquad\qquad$\includegraphics[angle=360,scale=0.6]{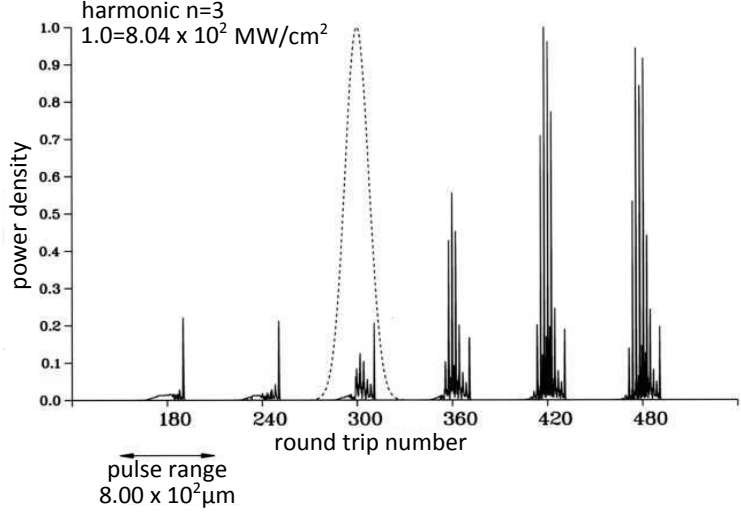}

\protect\caption{\label{Fig5}Pulse distribution of the third harmonic at different
round trips, same parameters as in Fig.\ref{Fig4}}

\end{figure}

\begin{figure}[H]
\includegraphics[angle=360,scale=0.5]{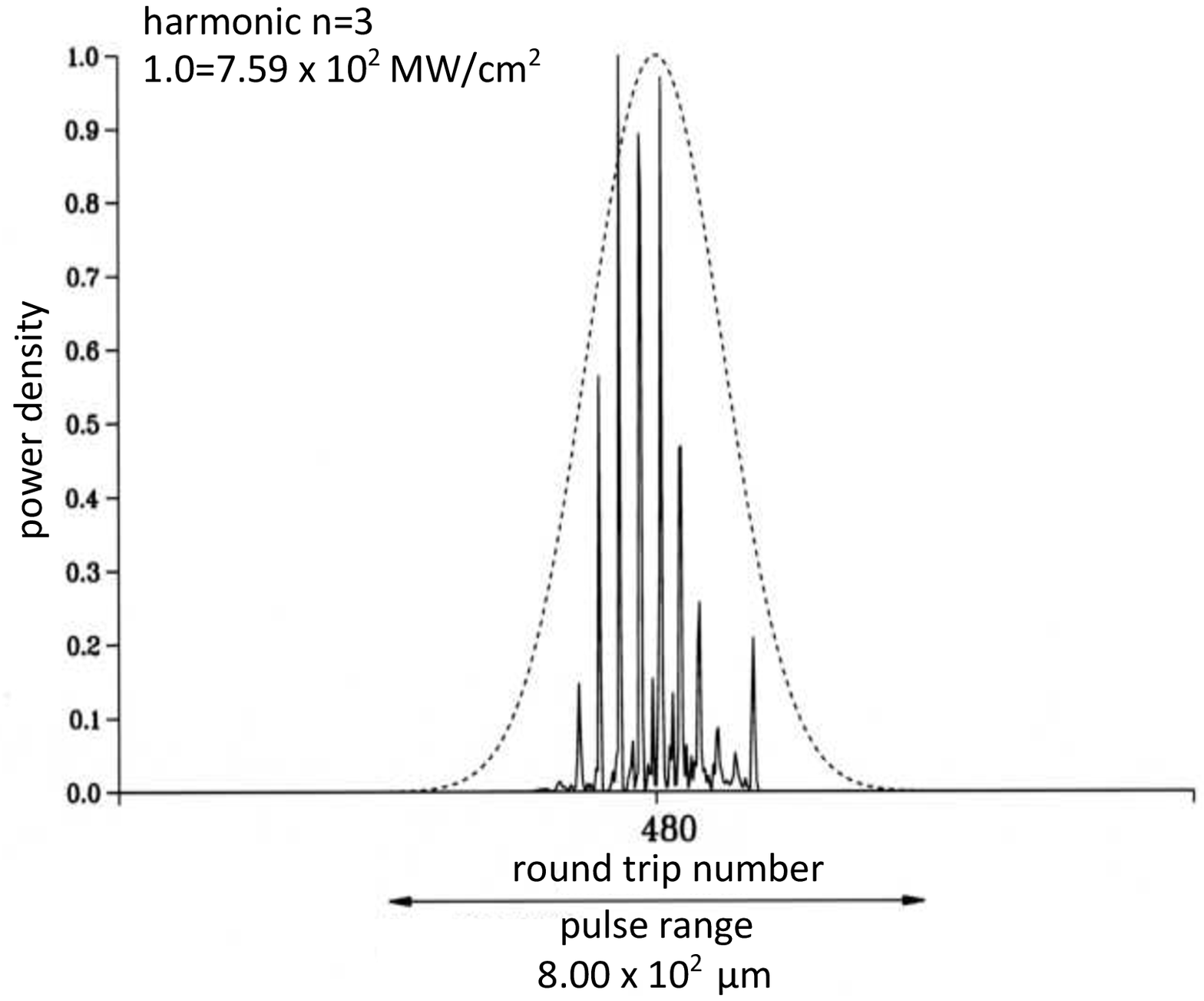}\includegraphics[angle=360,scale=0.49]{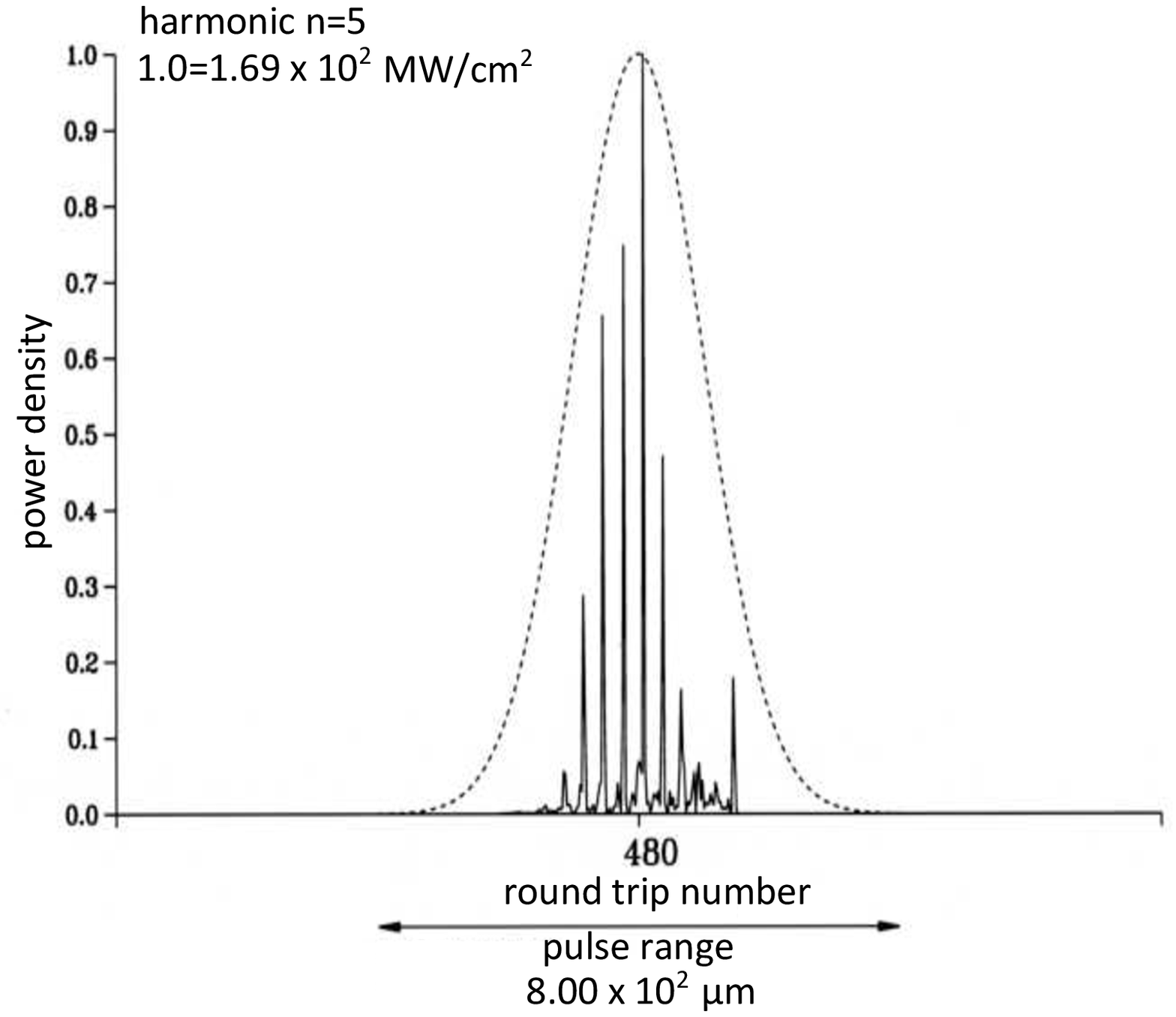}

\protect\caption{\label{fig:Fig6}Third and fifth harmonic distribution at 480-th round
trip, same parameters as in Fig.\ref{Fig4}}
\end{figure}

An interesting feature we observe in deep saturation is the presence
of a reinforcement of the power emitted at higher harmonics. This
effect should be understood as an increase in the efficiency of the
harmonic generation associated with the intrinsic super-radiant nature
of the emission process. The spiking distribution of the radiation
explores fresh region of the electron bunch ensuring a more copious
emission of radiation. An idea of the interplay between the pulse
distribution of the harmonic and the side band occurring at the fundamental
is provided in Figs. \ref{Fig5} and \ref{fig:Fig6}, where we have
reported both the pulse harmonic evolution at different round trip
and the “stationary” distribution 

The dynamics of the harmonic pulses, along with the relevant side
band, is quite transparent. Initially a more substantive harmonic
generation occurs on the trailing part of the beam where the bunching
is stronger, when the peaks shift towards the rear edge of the packet,
the side bands become increasingly larger in that region.

What we have described so far is the crude “phenomenology”, which
shows that the generation of stable sub-pulses with a length of the
order of a coherence length occurs in deeply saturated FEL oscillators.
The results we have so far obtained confirm the formation of side
bands at zero cavity detuning in deeply saturated conditions \cite{Nishimori,Hajima}
The existence of spikes in high gain FEL oscillators at zero desynchronism
has already been discussed in the past either theoretically and experimentally
in refs. \cite{Nishimori,Bonifacio} where the possible link with
the spikes characterizing the high gain behaviour has also been pointed
out. We have in addition pointed out that the pulses associated with
the higher order harmonics exhibit an analogous side band structure.
In the forthcoming section we will draw further consequences also
of practical nature.

\section{FEL deep saturated regime and output coupling}

We have stressed that the harmonic radiation is not stored in the
optical cavity but it represents just the power radiated on higher
harmonics at each round trip. The question arises therefore of how
the associated amount of power can be coupled out. Since in our example
we have considered a FEL operating in the visible (around 500 nm),
the third and fifth harmonic wavelength are at 165 and 100 nm, and
therefore much of the radiation could be absorbed by the mirrors.
This drawback becomes more significant at shorter wavelength. 

\begin{figure}
\includegraphics[scale=0.5]{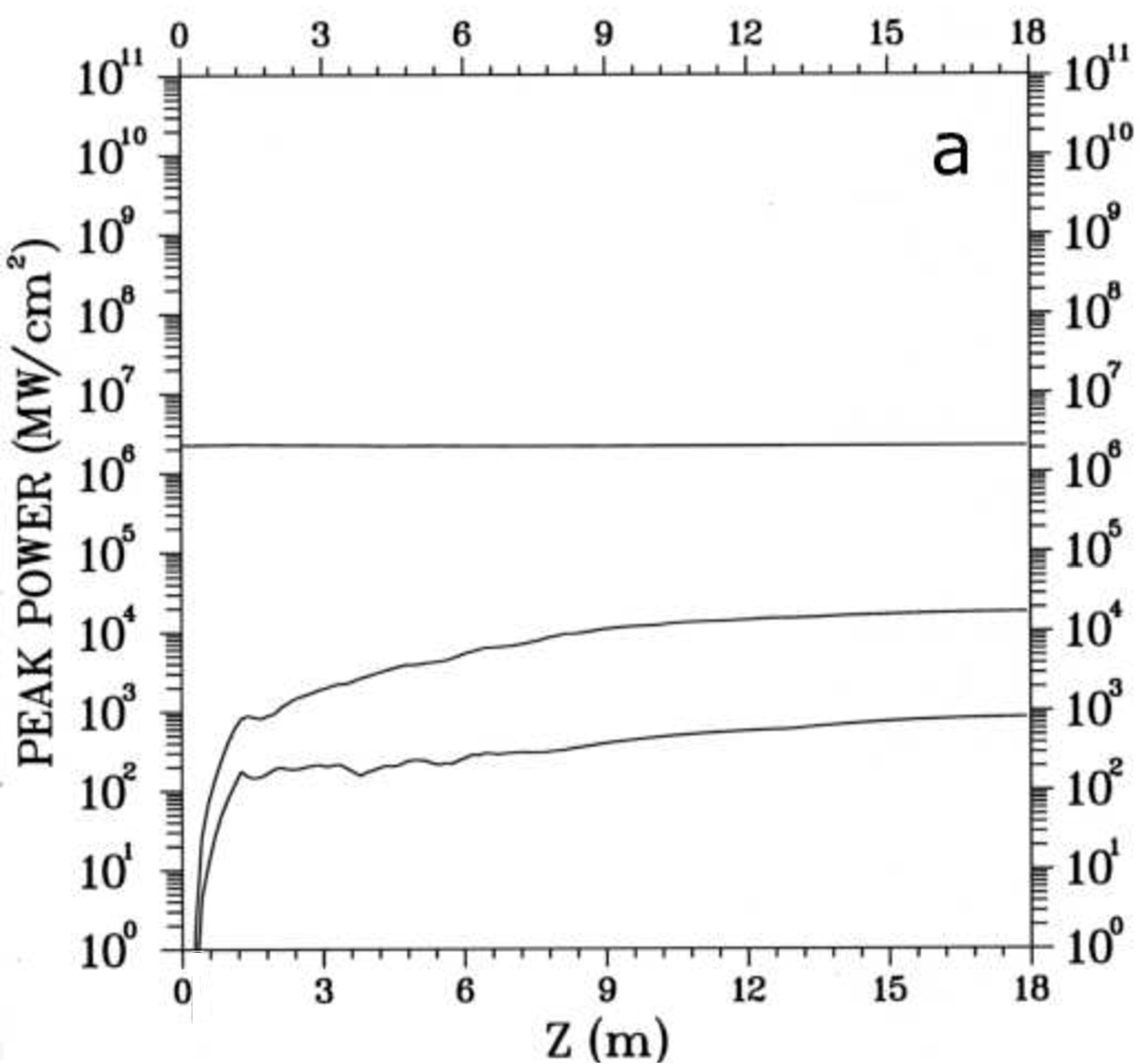}\enskip{}\includegraphics[scale=0.35]{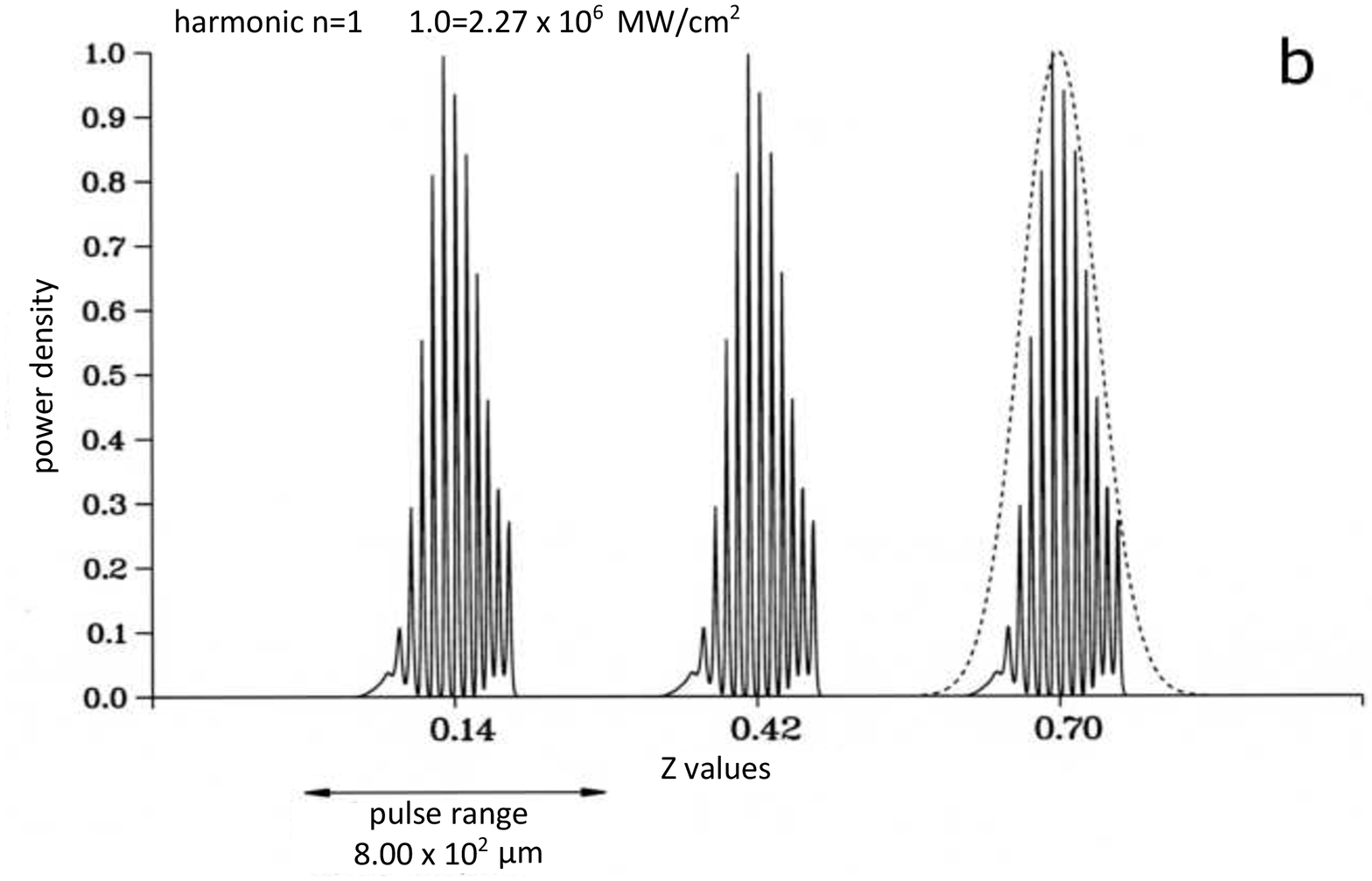}\includegraphics[scale=0.35]{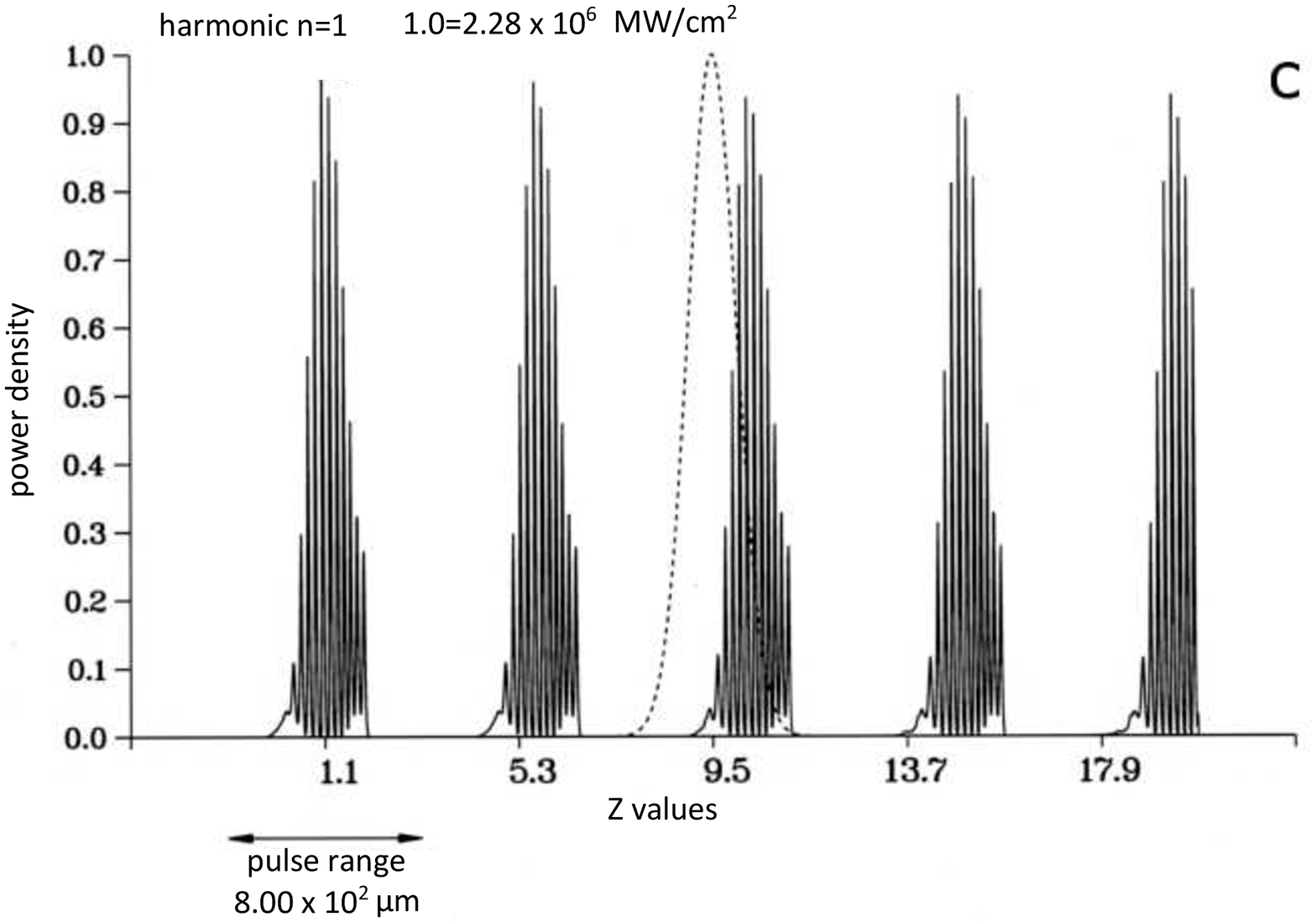}\enskip{}\includegraphics[scale=0.4]{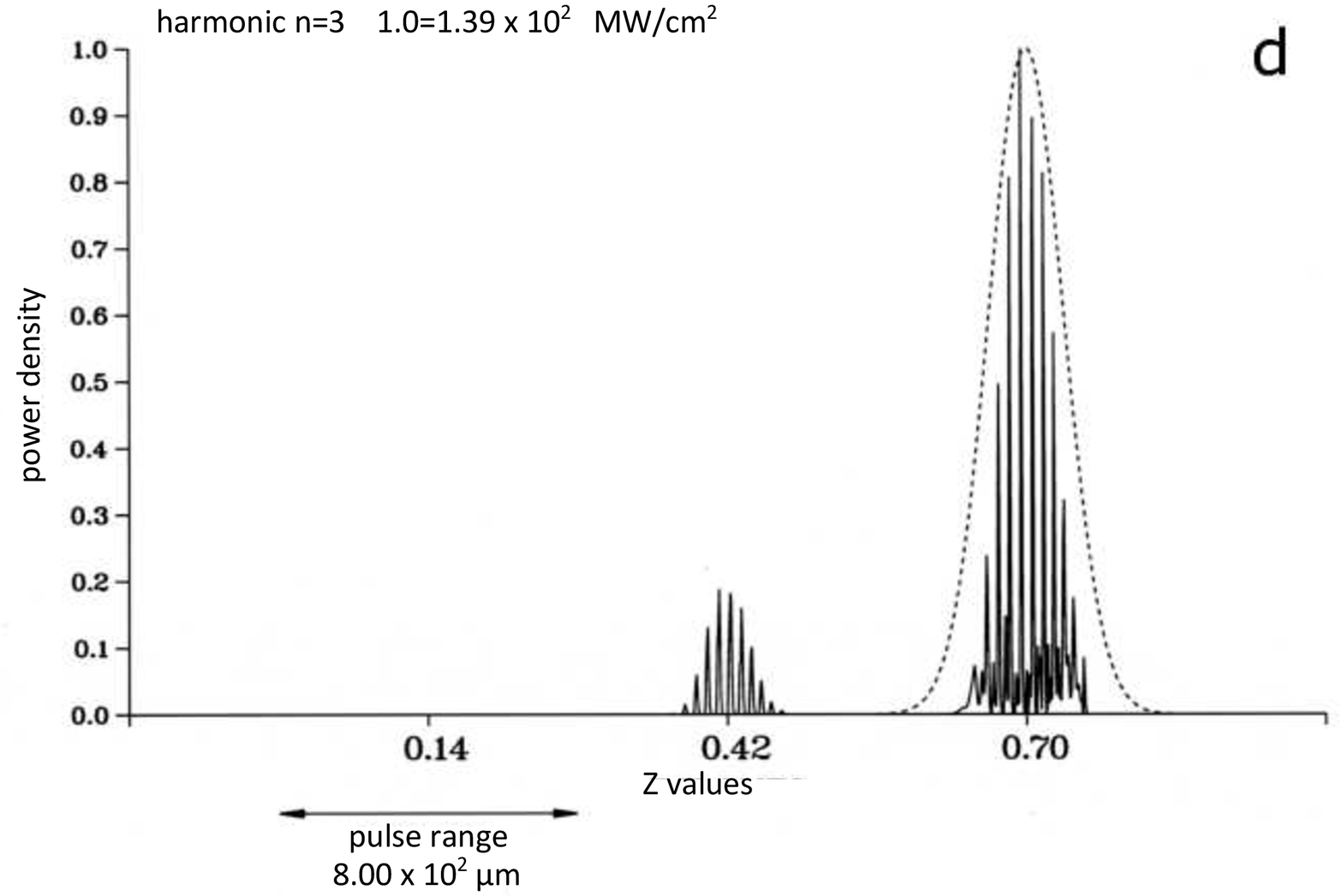}\includegraphics[scale=0.4]{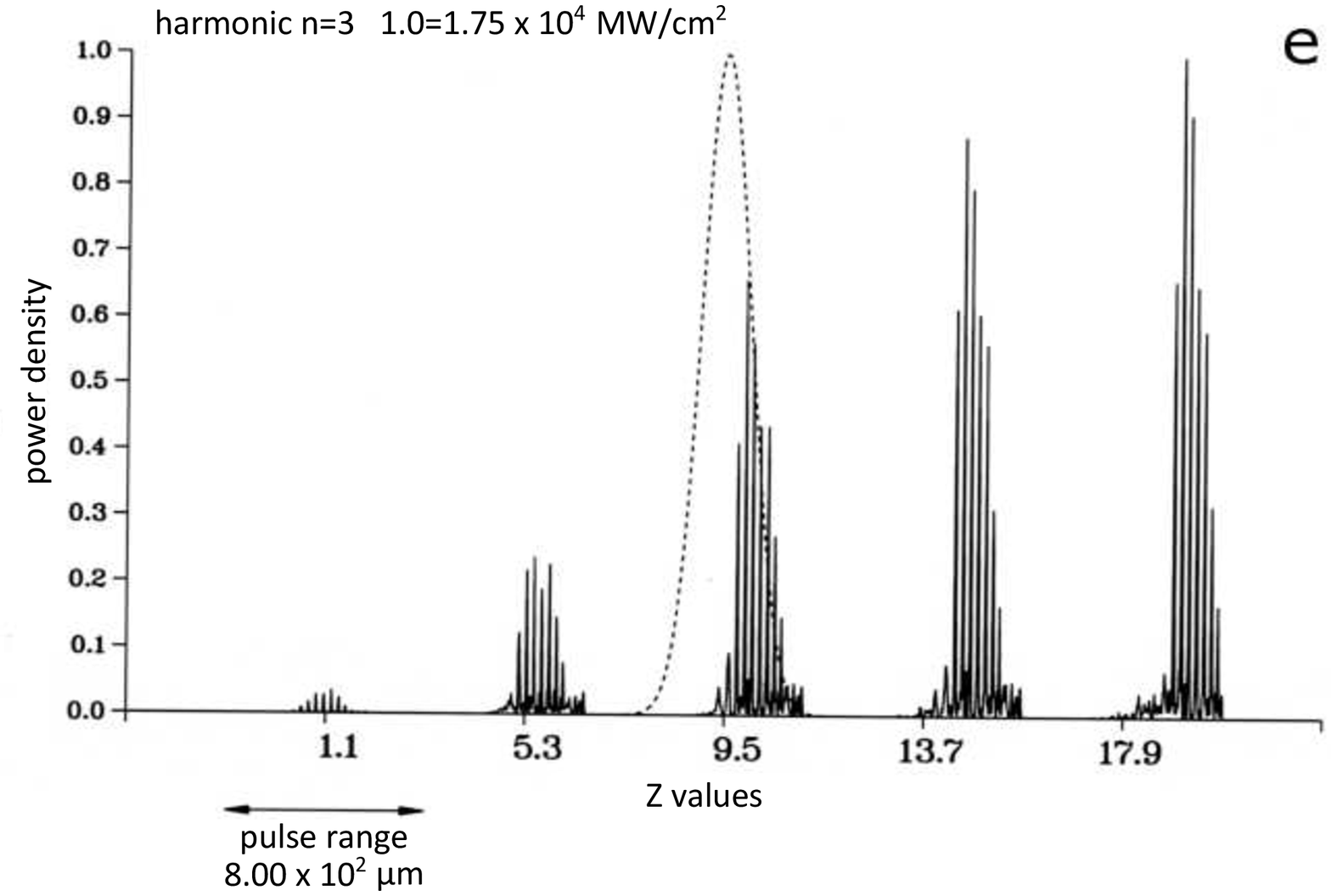}

\protect\caption{\label{Fig7}Radiation emitted by the electrons in the external undulator:
a) intensity growth for the first harmonic (which remains constant)
and third and fifth, the power level of the harmonics reaches the
maximum with a relatively short undulator; b) fundamental harmonic
in the first 0.7 m of the undulator; c) same as b up to 18 m; d) third
harmonic in the first 0.7 m of the undulator; e) same as d up to 18
m. }
\end{figure}

A problem of practical interest is therefore that of coupling the
harmonic power outside. A possibility is offered by the use of an
external radiator, provided by an undulator with the same characteristics
of the intracavity magnet. The e-b extracted from the cavity and injected
inside the radiator produces the effects shown in Fig. \ref{Fig7}.
When the oscillator operates in the deep saturated regime, the electrons
in the second undulator radiate an almost exact replica of the of
the intracavity first harmonic field intensity (albeit the associated
pulse slips over the electrons and acquires a more asymmetric shape
of its intracavity counterpart). The corresponding third harmonic
power grows very nicely and may reach significant power levels. It
is however remarkable that the level of laser power is so strong that
it impresses complete memory of its comb structure on the fresh electron
bunch, which is bunched in such a way to restore, in a kind of eco,
the harmonic generation inside the cavity. If this effect holds in
an actual experimental configuration, it could be exploited as an
out-coupler mechanism, thus providing an extremely useful extraction
tool, if e.g. the cavity optics absorb the radiation emitted at higher
harmonics. 

\begin{figure}
\includegraphics[scale=0.4]{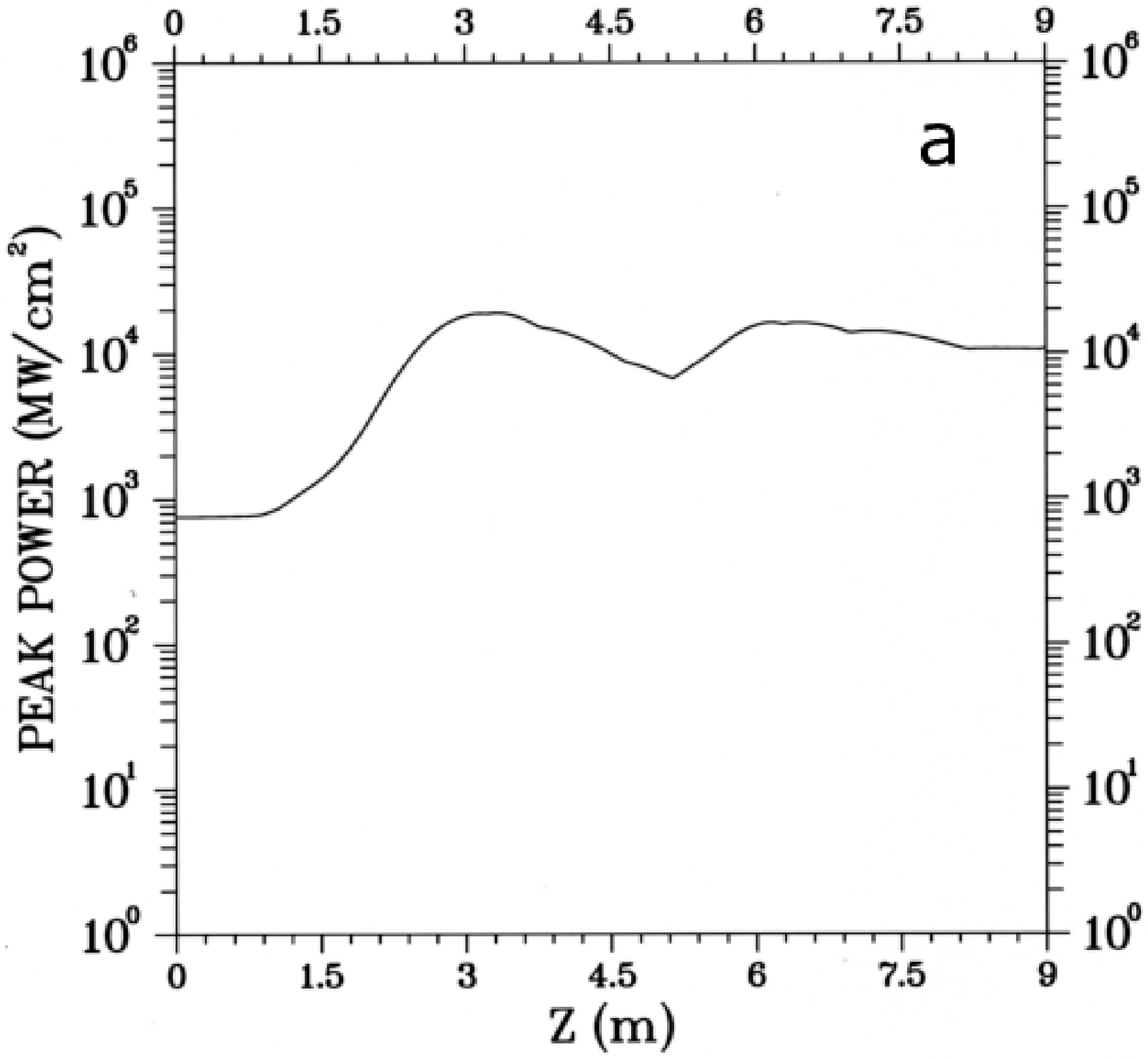}\enskip{}\includegraphics[scale=0.4]{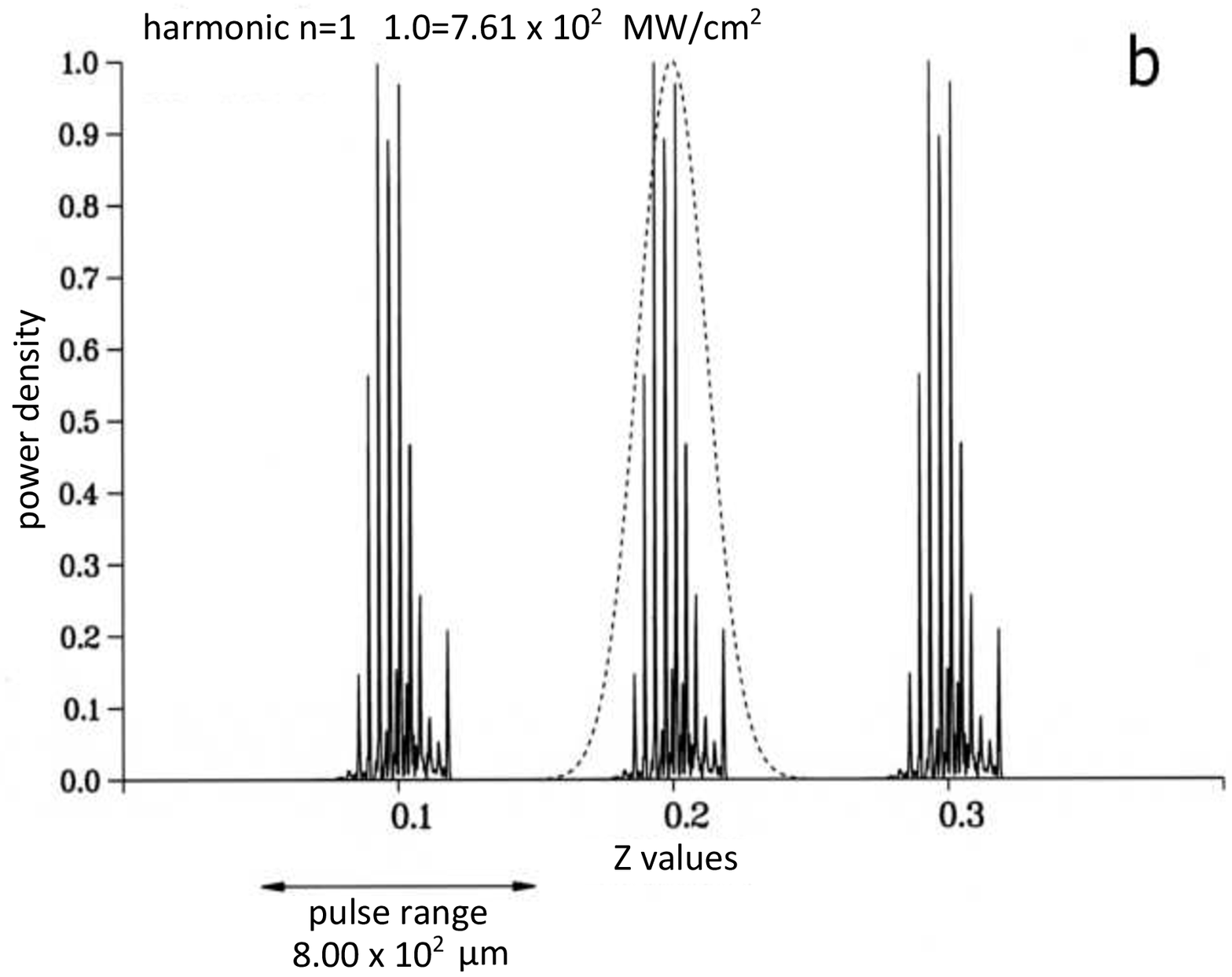}\includegraphics[scale=0.4]{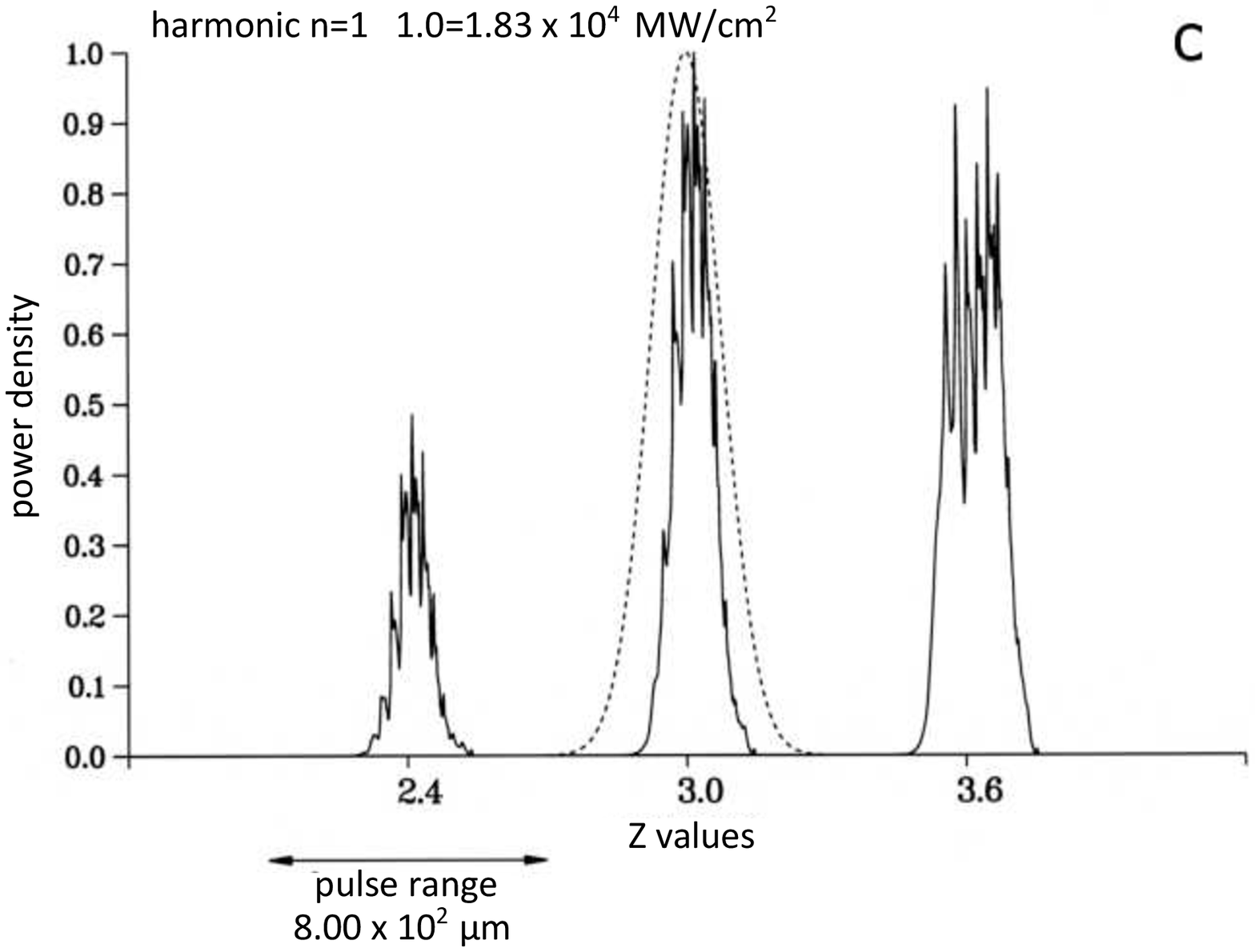}\enskip{}\includegraphics[scale=0.4]{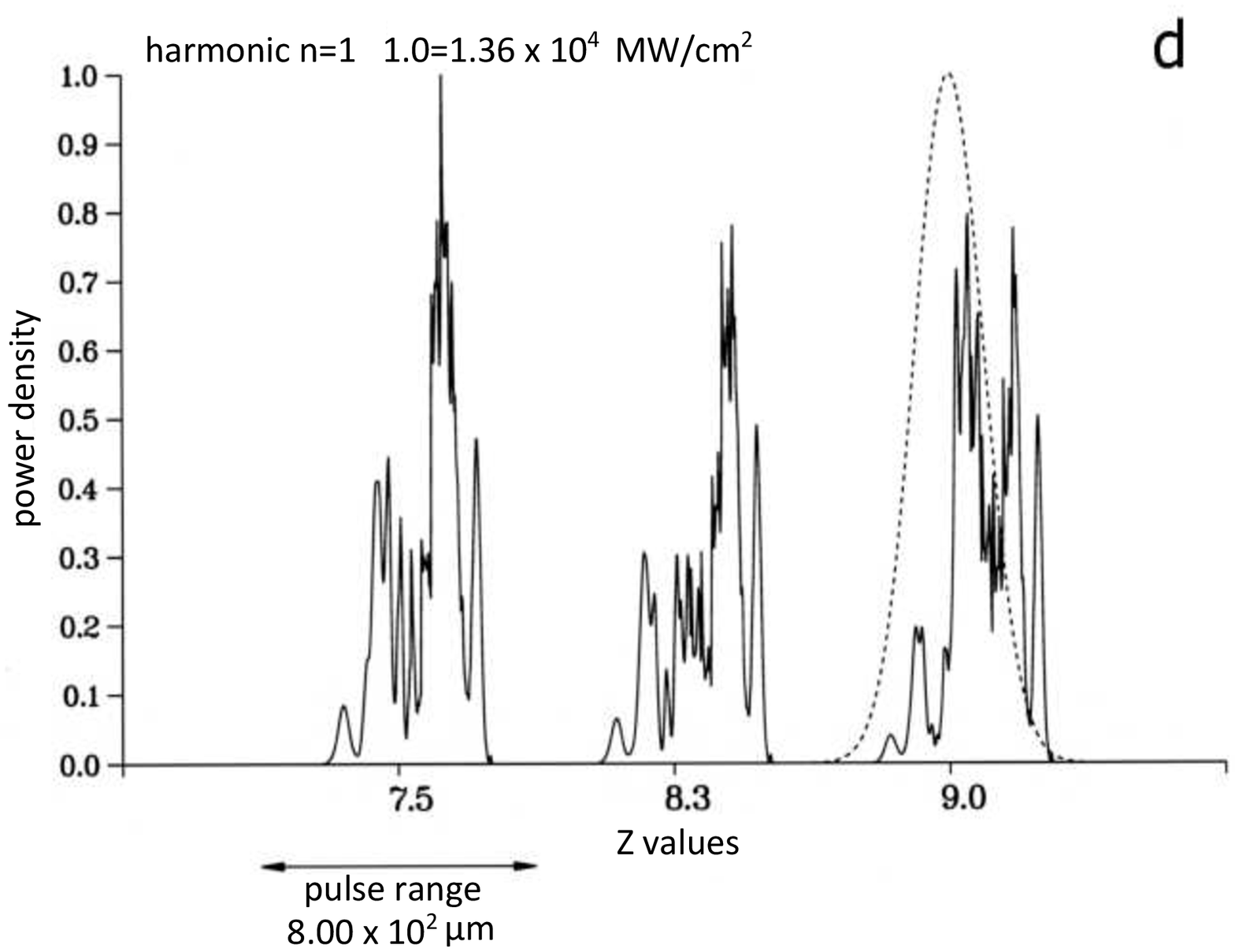}

\protect\caption{\label{Fig8}a) Harmonic power evolution outside the cavity; b) optical
pulse shape in the first undulator section (30 cm); c) Same as b)
further along the undulator; d) Same as b) along the last part of
the undulator .}

\end{figure}

A different situation occurs in the case in which the external undulator
is adjusted at the third harmonic of the first. In this case we obtain
what is indicated in Figs. \ref{Fig8}. The radiation is initially
extracted from the intra-cavity undulator in quite an unperturbed
form and is conserved over a small distance (of the order of one meter)
inside the external undulator. After this lethargic phase, the system
rearranges its phase space to adjust itself into a configuration which
has lost memory of the phase locking inside the oscillator and allows
intensity growth, with the unavoidable destruction of the comb structure. 

It is evident that the phase locking acquired in the cavity cannot
be maintained outside, if the external undulator has different characteristics
from that driving the oscillator. 

\section{Final Comments}

We have already guessed that the sub-peaks appearing in the deep saturated
oscillator regime could be viewed as the stable version of the spikes
occurring in the SASE operation. 

The intensity random fluctuations affecting the case of the latter
regime are not present in the oscillator configuration, because they
are locked in phase by the interplay of the intensity growth and by
the coupling of the longitudinal modes due to the electron bunch itself. 

The mechanism underlying the longitudinal mode coupling in pulsed
FEL oscillators has been discussed in the past and it was shown that
FEL multimode gain depends on the Fourier transform of the electron
beam itself \cite{DattCimento}. Such a phase locking provides the
first step towards the comb structure in deep saturated regime. 

The relevant dynamics can however been explained using as paradigmatic
well known mechanisms widely studied in conventional laser Physics. 

We note first that the intra-cavity evolution in the small signal
regime and for small cavity losses $\eta\ll1$ is dominated by a mechanism
analogous to the active mode locking by loss modulation \cite{Kaertner}.
In the present case the modulation occurs through the bunch current
shape. The relevant master equation for the complex field amplitude
A can be written as \cite{Dattoli-Renieri} 

\begin{eqnarray}
 &  & T_{R}\frac{\partial}{\partial T}A=\left[\left(G_{1}-\eta\right)+\mu_{c}\left(G_{2}-\theta\right)\frac{\partial}{\partial\tau}+\frac{1}{2}G_{3}\frac{\partial^{2}}{\partial\tau^{2}}-\frac{1}{2}G_{1}\tau^{2}\right]A\nonumber \\
 &  & \tau=\frac{z}{\sigma_{z}},\;T_{R}=2\frac{L_{c}}{c},\;L_{c}\equiv Cavity\:length\label{eq:Eq7}
\end{eqnarray}

where $\tau$ is the longitudinal coordinate, normalized to the bunch
rms length, $G_{1,2,3}$ are complex quantities associated to the
FEL gain and $\eta$ and $\theta$ have been defined previously. Furthermore,
the equation has been obtained by approximating the finite difference
equation, in terms of round trip number, with a first order differential
equation in the time $T$ defined as $T\approxeq rT_{R}$.

Eq. (7) holds for optical pulses around the centre of the electron
bunch and is fully equivalent to its counterpart in conventional laser
physics. There are several reasons why we have reported eq. (\ref{eq:Eq7}),
which holds in the case of small signal low gain regime. Apart from
the fact that there is a one to one mapping with symilar equations
accounting for active mode locking in conventional lasers , we note
that from the mathematical point of view, eq. (\ref{eq:Eq7}) is a
master equation with a quadratic potential. Its stationary solutions
are provided by Hermite-Gauss modes, and provide the Super Modes for
the problem under study \cite{Ellaume,dattoli-delfra}, namely small
slippage compared to the electron bunch length. The gain of each eigenmode
is not zero at 0 cavity detuning (the maximum gain is given by the
condition $\theta=G_{3}$ ). Equation (7) can be used to model the
saturation provided that the gain terms be replaced by the corresponding
gain saturated forms (see Appendix for further details). It should
be stressed that eq. (7) assumes that the mode growth occurs around
the maximum of the electron bunch distribution and indeed the quadratic
terms in $\tau^{2}$ is due to an expansion of the current profile
up to the lowest order in the longitudinal coordinate. This is also
compatible with the assumption that the slippage is small compared
to the bunch length and, therefore, that the mode longitudinal extension
cover a limited portion of the electron bunch. The small slippage
approximation allows the possibility of studying the field evolution
around the center of the electron packet. Other SM like forms may
grow in correspondence of other positions inside the e-bunch, after
saturation around the maximum of the current has occurred. 

The fundamental mode, namely that corresponding to the lowest order
eigenmode of eq. (\ref{eq:Eq7}) , has the largest gain and is the
more likely to reach saturation, it has a gaussian shape, having the
width reported in eq. (\ref{eq:ec.6}) (See Appendix). 

We must however emphasize that during the evolution the optical pulse
tends to move towards the front end of the electron bunch, this determines
a faster saturation of the “active-medium” represented by this portion
of the bunch. It should now be kept in mind that the lowest order
SM is that reaching the saturation, while the others die, before that
the system approaches the equilibrium. The survival of only one mode
ensures the stability, unless other factors like beam jitter or cavity
length fluctuations are present. As already remarked the SM equation
(7) is appropriate for the description of the laser evolution around
the maximum of the electron current, which is the center of the optical
packet. On the other side, in view of the fact that the radiation
explores only a small portion of the electron bunch during its passage
inside the undulator, other SM-like structure may develop if sufficient
gain is available around different regions of the electron packet.
They are ruled by equations analogous to those given in (7), apart
from a gain modulation determined by the current profile. If we sample
the electron bunch with the number of peaks \textquotedbl{}predicted\textquotedbl{}
by eq. (6), we obtain (numerically) what has been described in the
previous parts of the paper, namely the formation of narrow subpeaks,
which agrees (at least in number) with the fluctuating spikes of the
SASE regime. It should also be noted that they have a spatial width
of the same order of the coherence length. This is what we have found
numerically using Prometeo code and an attempt of theoretical justification
is provided in the Appendix.

The previous description is certainly not sufficient to explain the
formation of the deep saturated comb like structure in oscillators
and further remarks are in order, mainly in connection with the packet
shortening. As in conventional laser devices saturation is driven
by a decrease of the gain due to the intra-cavity intensity through
a factor of the type 

\begin{eqnarray}
 &  & G\left(T\right)\propto\frac{G}{1+\alpha I},\nonumber \\
 &  & I\propto|A|^{2}\label{eq:Eq8}
\end{eqnarray}

Furthermore, since the pulse is travelling in a medium, the Kerr effect
\cite{Kaertner} produces a change of the refractive index with the
intensity, namely 

\begin{equation}
n\approxeq n_{0}+n_{2}I\label{eq:Eq9}
\end{equation}

where $n_{0},\:n_{2}$ are the linear and the second order non-linear
refractive indices characterizing the FEL medium. This effect produces
a shift in the instantaneous phase of the pulse which in turns determines
a frequency shift, leading to a frequency chirping inducing a compression
of the pulse. 

The mechanism is slightly more complex than the effect on single pulses
propagating in selfoc fibers due to the appearance of multiple peaks,
which is simply due to the finite size of the coherence length, which,
in the case of our simulations, is much shorter than the electron
bunch length. An insightful analysis on this point can also be found
in \cite{Haiima-Nagai}.

\begin{figure}[H]
$\qquad\qquad\qquad\qquad$\includegraphics[bb=0bp 0bp 554bp 770bp,scale=0.42]{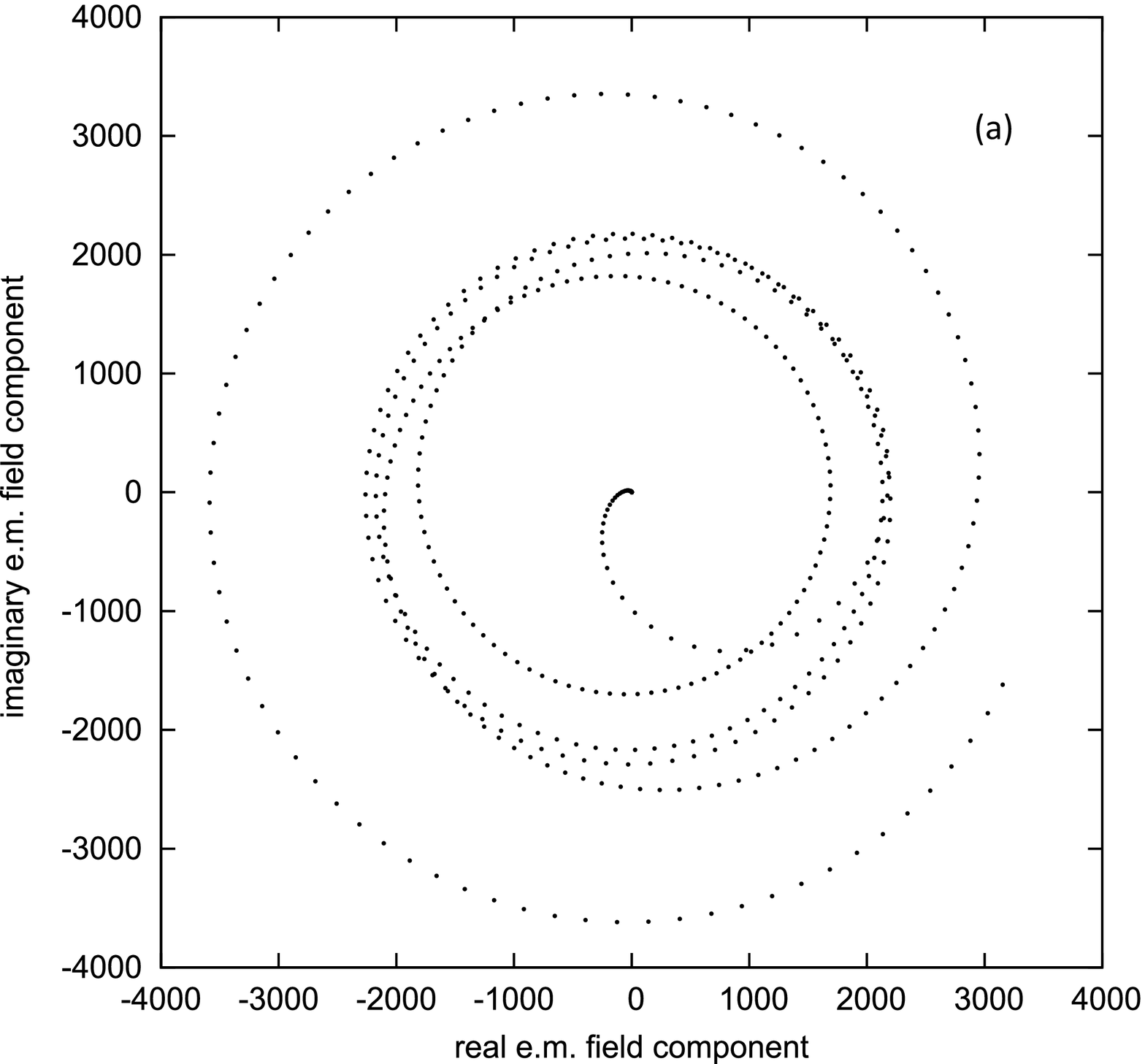}

$\:\qquad\qquad\qquad\qquad$\includegraphics[scale=0.42]{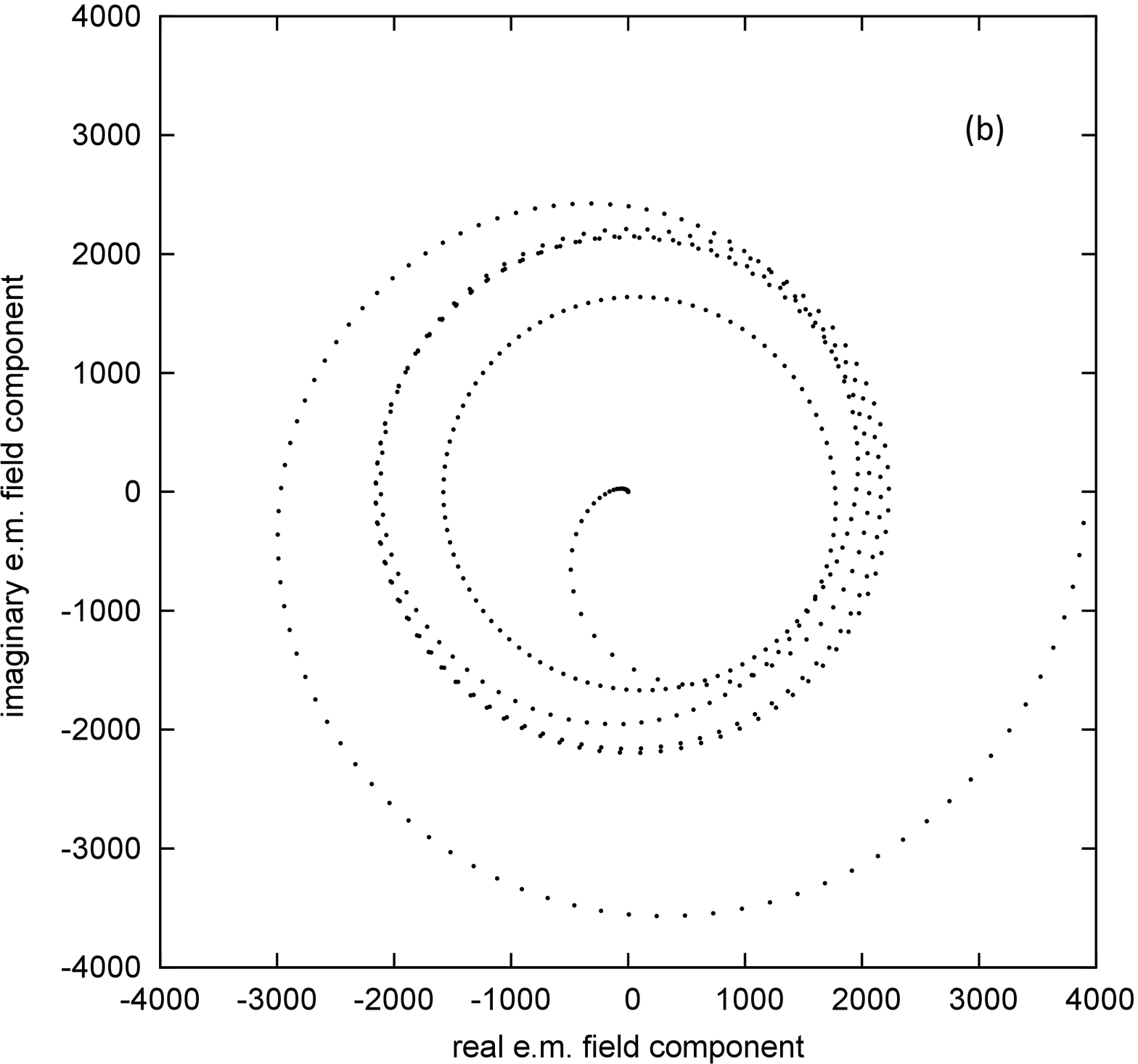}

\protect\caption{\label{fig:Fig9}Round trip evolution of the real (abscissa) and imaginary
(ordinate) parts of the fields in correspondence of the maximum of
the sub peaks in Fig.\ref{Fig.3}: a) peak 7; b) peak 6. (It is to
be noted that peak 7 approaches a stationary configuration ).}

\end{figure}

This is just a qualitative statement, a more detailed study in analytical
terms will be presented elsewhere. A further idea of the sub-peak
evolution is provided by Fig. \ref{fig:Fig9} where we have reported
the Nyquist diagram of a pair of different peaks. The plots provide
the round trip evolution of the complex vector associated with the
relevant field amplitude. We have chosen the case of the spikes 6
and 7.

In fig. 10a) we have reported the cosine of the phase difference of
peaks 6 and 7 vs. the round trip number, for rt larger than 200 when
the comb structure is being evidentiated. The phase difference in
this region is reasonably constant for a large value of rt and the
peaks can reasonably considered phase locked. In Fig. 10b) we have
taken distant spikes (1-10) and the phase locking condition is reached
in longer times.

\begin{figure}[H]
$\qquad\qquad\qquad\qquad$\includegraphics[scale=0.4]{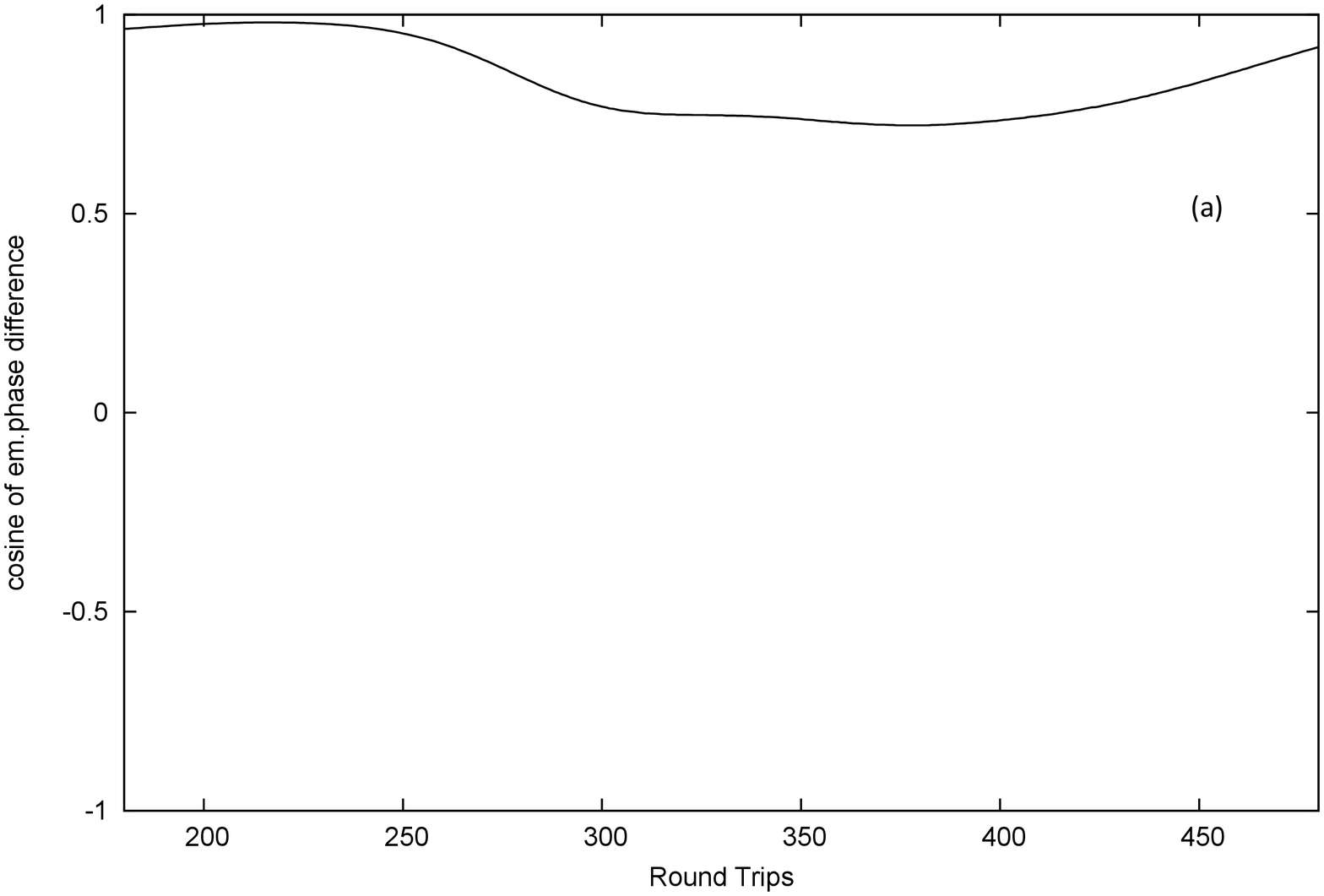}

$\:\qquad\qquad\qquad\qquad$\includegraphics[scale=0.4]{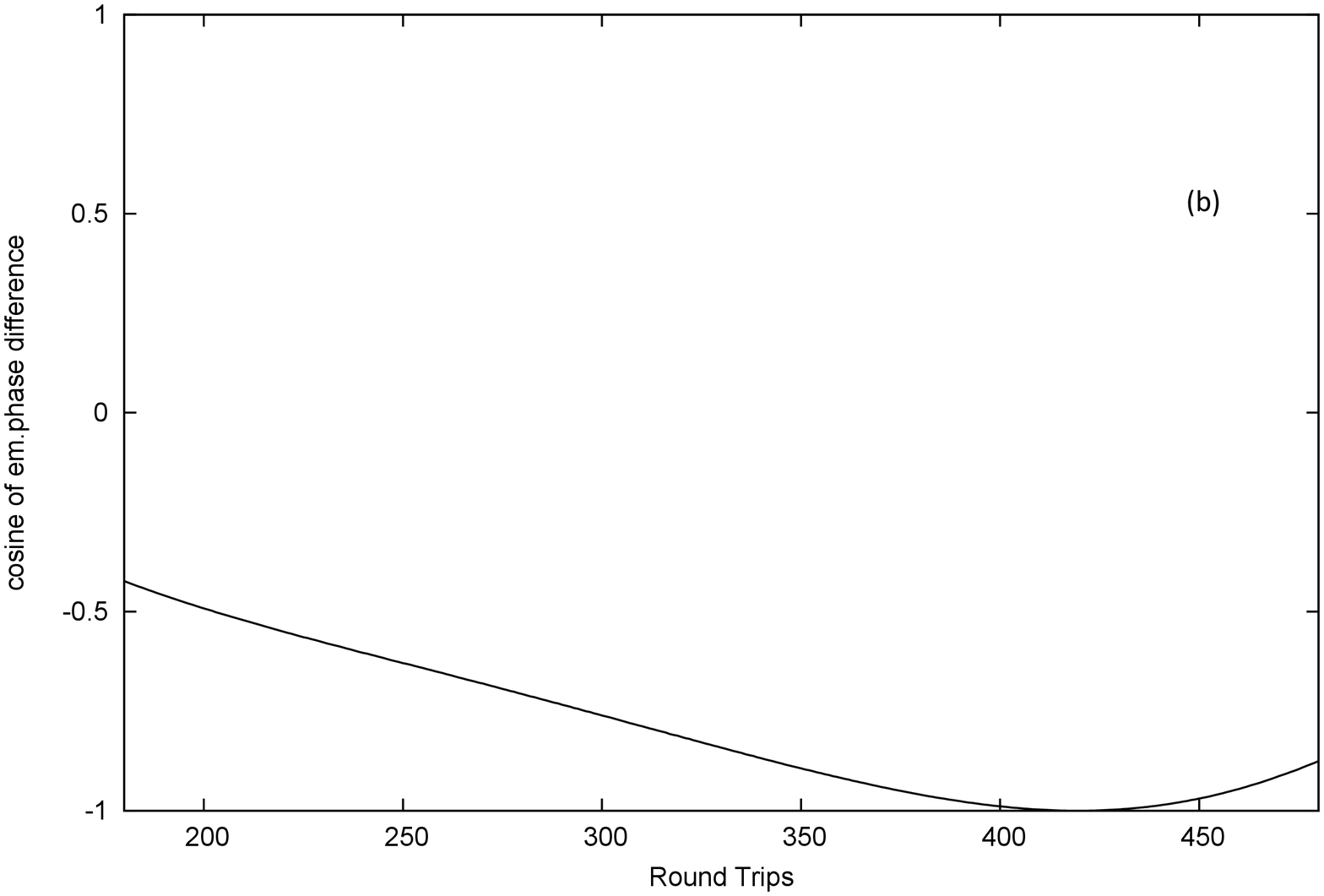}

\protect\caption{Round trip evolution of the cosine of the phase difference of the
spikes 6 and 7 (a) and the spikes 1 and 10 (b)}

\end{figure}

We have stressed that the peak occurs at zero detuning, when the cavity
is differently settled the formation of the peaks is strongly reduced.
We have run the code for different values of the detuning and verified
the absence of spikes deep saturated comb-like structures for $\theta\neq0$
. 

A final point we want to stress is the comparison of the present result
with a full 3-D code, to this aim we have modified GENESIS to run
it in the oscillator mode. In Fig. 11 we have reported the field distribution
at different round trip, along with the relevant spectra. The pattern
of the peak formation is qualitatively similar to the dynamics predicted
by the 1-D dynamics. We finally stress the content of Fig. 12, reporting
the spectrum distribution at a very large round trip number ($r=800$):
the result is impressive, the spectral structure is that of a SASE
run, the only (and significant) difference is that the structure is
stationary, namely it is self-reproducing after each round trip. 

In this paper we have touched, with some detail, the mechanism of
“spiking mode locking” in deeply saturated oscillators. Most of our
considerations and results find eco in previous research \cite{Haiima-Nagai}
where the problem of the production of stable attosecond pulses has
been discussed in depth. We want however to make a further remark
on the results of the paper which can be summarized as the possibility
of observing a comb structre in the laser pulses and on the relevant
feedback on the non linear harmonic generation. The possibility of
observing these effects are associated with the use of accelerating
devices using e-beam with long macropulses (like Super conduting Linacs)
and on the use of smooth outcoupling magnet allowing the injection
of the e-beam in the second undulator \cite{Kulipanov}. In this case
particular care should be devoted to design the magnets to avoid coherent
synchrotron radiation emission, which might prevent the generation
of higher order harmonics with the associated comb structure. A more
detailed discussion will be presented in a forthcoming investigation,
in which we will discuss how these effects can be implemented on the
IRIDE-FEL architecture \cite{IRIDE}.

In a forthcoming investigation we will derive the mode locked slice
structure by the use of a theory based on the extension of the technique
of passive mode locking of conventional laser physics. 

\begin{figure}[H]
\includegraphics[scale=0.65]{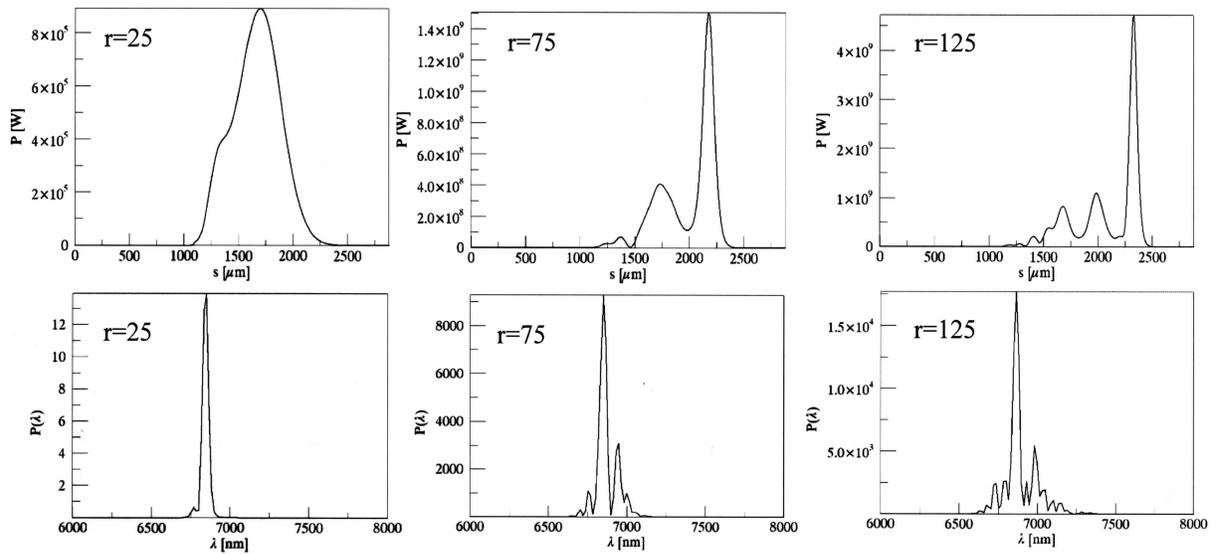}

\protect\caption{\label{fig:Fig11}First line: pulse power vs position s for different
values of trip number r. Second line: Spectral Intensity vs wavelength.
Simulations by GENESIS 1.3. }
\end{figure}

\begin{figure}[H]
\includegraphics[scale=0.5]{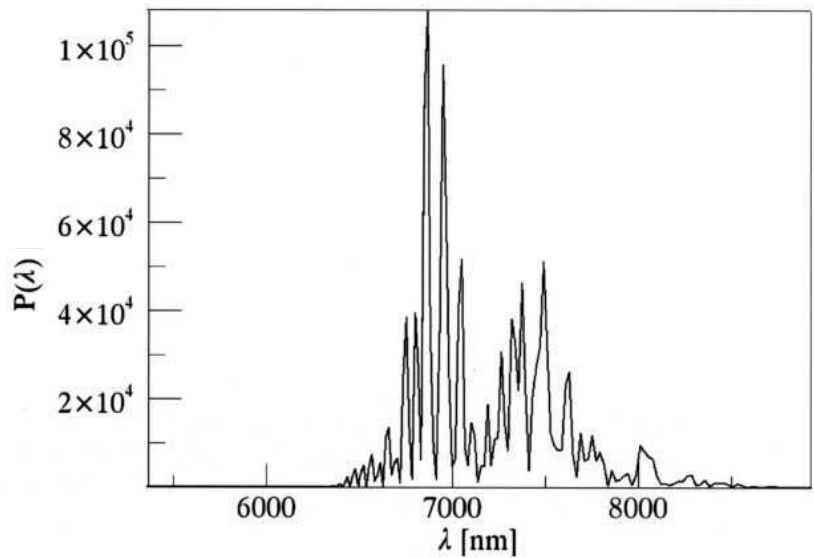}

\protect\caption{\label{fig:Fig12}Spectral intensity vs wavelength at r=800. Simulations
by GENESIS 1.3.}
\end{figure}

\bigskip{}

\textbf{\Large{}Appendix}{\Large \par}

\bigskip{}

In this Appendix we will discuss more quantitatively some points previously
touched in qualitative terms only. 

We will first clarify how the coupling parameter $\mu_{c}$ emerges
quite naturally in the analysis of FEL gain. We assume that the electron
bunch exhibits the Gaussian longitudinal distribution

\begin{equation}
f\left(z\right)=\frac{1}{\sqrt{2\pi}\sigma_{z}}e^{-\frac{z^{2}}{2\sigma_{z}^{2}}}\label{eq:A1}
\end{equation}

The relevant Fourier Transform (FT) writes

\begin{equation}
\tilde{f}\left(k\right)=\frac{1}{2\pi\sigma_{z}}\intop_{-\infty}^{\infty}e^{-i\left(k-k_{s}\right)z}e^{-\frac{z^{2}}{2\sigma_{z}^{2}}}dz.\label{eq:A2}
\end{equation}

We have denoted by $k$ the wave vector of a longitudinal mode. The
role of the bunch distribution is that of filtering the longitudinal
modes around the resonant frequency ( $\omega_{s}=ck_{s}$ ). Using
therefore as reference parameter the frequency detuning

\begin{equation}
\nu=2\pi N\frac{k-k_{s}}{k_{s}}=2\pi N\frac{\omega_{s}-\omega}{\omega_{s}}\label{eq:A3}
\end{equation}

we can write the FT in the form 

\begin{equation}
\tilde{f}\left(\nu\right)=\frac{1}{2\pi\sigma_{z}}\intop_{-\infty}^{\infty}e^{-i\frac{\nu}{N}\frac{z}{\lambda_{s}}}e^{-\frac{z^{2}}{2\sigma_{z}^{2}}}dz=\frac{1}{\sqrt{2\pi}}e^{-\frac{\nu^{2}}{2\mu_{c}^{2}}}\label{eq:A4}
\end{equation}

The effective FEL gain accounting for the longitudinal distribution
of the current is accordingly obtained by a convolution of the ordinary
gain function on the distribution function (\ref{eq:A4}). 

The FEL gain itself is obtained by making a convolution on the gain
bandwidth with the electron distribution, things should however been
done by including also the effect of the slippage. The small signal
field amplitude evolution in one undulator passage can be written
as

\begin{equation}
\begin{array}{c}
\frac{d}{d\bar{t}}a=i\pi\sqrt{2\pi}g_{0}f\left(z+\Delta\cdot\bar{t}\right)\left[\intop_{0}^{\bar{t}}d\bar{t'}\left(\bar{t}-\bar{t'}\right)e^{-i\nu\left(\bar{t}-\bar{t'}\right)}a\left(z+\Delta\cdot\bar{t'},\bar{t'}\right)\right],\\
\bar{t}=\frac{z}{L_{u}},\:L_{u}\equiv undulator\:length
\end{array}\label{eq:A5}
\end{equation}

After expanding either the current and the field up to the second
order in the slippage parameter $\Delta$ we obtain the (small signal)
pulse propagation equation valid for $\Delta\ll\sigma_{z}$ \cite{dattoli-delfra}.
The master FEL equation provided by eq. (\ref{eq:Eq7}) is obtained
from (\ref{eq:A5}) after performing the low gain approximation $a\left(\bar{t'}\right)\cong a\left(\bar{t}\right)$
and iterating the single pass solution to an arbitrary number of round
trips in an optical cavity, the gain equations appearing in the master
equations are just given by 

\begin{equation}
\begin{array}{c}
G_{2}\propto i\partial_{\nu}G_{1}\\
G_{3}\propto-\partial_{\nu}^{2}G_{1}
\end{array}\label{eq:A6}
\end{equation}

And $G_{1}$ is the complex gain function \cite{Dattoli-Renieri}

\begin{equation}
G_{1}=\frac{\pi}{\nu^{3}}\left[2\left(1-e^{-i\nu\tau}\right)-i\nu\tau\left(e^{-i\nu\tau}+1\right)\right].\label{eq:A6a}
\end{equation}

We have denoted in eq. (\ref{eq:Eq7}) the field amplitude by “$A$
“ and not by “$a$ “ to stress the difference between the single pass
and oscillator master equation. Most of this procedure and the relevant
link with conventional laser Physics \cite{Kaertner} has been discussed
in \cite{dattoli-delfra}.

Note that we have 

\begin{equation}
\begin{array}{c}
A_{n}\left(\tau,T\right)=\alpha_{n}\left(\tau\right)e^{\lambda_{n}\frac{T}{T_{R}}},\\
\alpha_{n}\left(\tau\right)\propto\frac{1}{\sqrt{2^{n}\sqrt{\pi}\tau_{0}}}H_{n}\left(\frac{\tau}{\tau_{0}}\right)e^{-\frac{\tau^{2}}{2\tau_{0}^{2}}},\\
\tau_{0}\cong\frac{1}{2}\sqrt{\mu_{c}},\:\lambda_{n}\cong G_{M}-\eta-\frac{2}{3}\mu_{c}\left(n+\frac{1}{2}\right)
\end{array}\label{eq:A6b}
\end{equation}

($G_{M}=$maximum small signal gain).

Using non normalized variables we find for the width of the SM

\begin{equation}
\sigma_{b}=\sigma_{z}\tau_{0}\cong\frac{1}{2}\sqrt{\Delta\sigma_{z}}.\label{eq:A7}
\end{equation}

It is evident that the lowest order eigenmode ($n=0$) is that with
largest gain. 

The use of the FEL gain saturation formulae yields 

\begin{equation}
\begin{array}{c}
\lambda_{n}\cong G_{M}\left(I_{n}\right)-\eta-\frac{2}{3}g_{0}\mu_{c}\left(n+\frac{1}{2}\right)\\
G\left(I_{n}\right)=G_{M}\frac{1-e^{-\beta X_{n}}}{\beta X_{n}},\:\:\:\beta=1.0145\cdot\frac{\pi}{2},\:\:X_{n}=\frac{I_{n}}{I_{s}}
\end{array}\label{eq:A8}
\end{equation}

Where the saturation intensity $I_{s}$ has been assumed to be the
same for all the Super Modes.

The intracavity equilibrium power for $\eta\ll1$ , can be evaluated
as (we refer to the fundamental SM)

\begin{equation}
I_{e}\cong\left(\sqrt{2}+1\right)\left(\sqrt{\frac{G_{M}}{\eta}}-1\right)I_{s}\label{eq:A10}
\end{equation}

The pendulum equation written in normalized variables writes 

\begin{equation}
\begin{array}{c}
\frac{d^{2}}{d\bar{t}^{2}}\varsigma=\left|a\right|\cos\left(\varsigma+\phi\right),\\
\left|a\right|^{2}\cong8\pi^{2}\frac{I}{I_{s}}
\end{array}\label{eq:A11}
\end{equation}

thus getting, at equilibrium

\begin{equation}
\left|a\right|\cong\sqrt{\left(\sqrt{2}+1\right)8\pi^{2}}\sqrt{\frac{1}{\sqrt{\frac{G_{M}}{\eta}}-1}}\cong13.8\sqrt[4]{\frac{\eta}{G_{M}}}\label{eq:A12}
\end{equation}

At saturation the side band instability develops and the associated
spikes correspond to half rotation of the electrons in the phase space,
the necessary time for this cycle is \footnote{This argument has been suggested by an anonymous Referee.}

\begin{equation}
\Delta T\cong\frac{L_{u}}{2\sqrt{2}c}\sqrt{\frac{I_{s}}{I}}\label{eq:A13a}
\end{equation}

and, at equilibrium, we find 

\begin{equation}
\Delta T\cong\frac{L_{u}}{4.4c}\sqrt[4]{\frac{\eta}{G_{M}}}\label{eq:A13b}
\end{equation}

The length of the coherent optical pulse associated with the side
band instability spike, could be inferred from the slippage length
occurring in one spiking cycle, namely

\begin{equation}
\sigma\cong\frac{N\lambda}{4.4}\sqrt[4]{\frac{\eta}{G_{M}}}\label{eq:A14}
\end{equation}

which, using the identity

\begin{equation}
N\cong\frac{\left(\pi g_{0}\right)^{\frac{1}{3}}}{4\pi\rho}\label{eq:A15}
\end{equation}

yields 

\begin{equation}
\sigma\cong l_{g}\sqrt{3}\frac{\left(\pi g_{0}\right)^{\frac{1}{3}}}{4.4}\sqrt[4]{\frac{\eta}{G_{M}}}\label{eq:A16}
\end{equation}

For all the reasonable values of a FEL oscillator we obtain

\begin{equation}
\sqrt{3}\frac{\left(\pi g_{0}\right)^{\frac{1}{3}}}{4.4}\sqrt[4]{\frac{\eta}{G_{M}}}\cong1,\label{eq:A17}
\end{equation}

the width of the side band spikes is therefore of the same order of
the coherence length.

We have considered the dependence of the widths of the comb peaks
on the cavity losses (see Fig. 13 ) and on the small signal gain coefficient
value. The plot are the result of a simulation relevant $g_{0}\cong0.3$
and losses $\eta\cong0.03$ accordingly we obtain a number of spikes
( $n_{s}=6$ ) consistent with eq. (\ref{eq:ec.6}) and with a width
comparable with $l_{c}$ . We have tried different combinations of
the parameters (gain losses, bunch length…) and the scaling relations
we have hypothesized have been confirmed, along with the fact that
the dependence of the sub pulse width on the cavity losses is weak.

\begin{figure}
\includegraphics[scale=0.8]{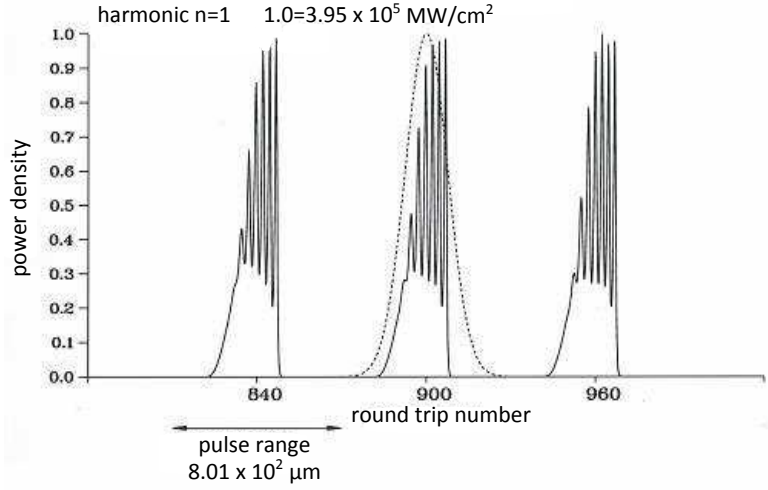}\protect\caption{FEL saturated pulses at different round trips, same parameters of
Fig. 2.}

\end{figure}

It is evident that the previous considerations do not corroborate
the suggestions that the saturated comb-spike are associated with
high gain spikes, but provide a further independent argument that
their duration coincides with the coherence length. 

We want further to stress that when saturation has occurred around
the maximum of optical packet, other regions may contribute to the
growth of the field itself and should be distributed according to
the “normal modes” associated with the oscillator FEL dynamics, Super
Modes provide one of these possible sets. We have therefore assumed
that the comb structure we have obtained in our numerical runs can
be obtained as 

\begin{equation}
\begin{array}{c}
S\left(z\right)\propto\alpha_{n_{s}}\left(z\right)e^{-\frac{z^{2}}{2\sigma_{z}^{2}}}\\
\alpha_{n_{s}}\cong\frac{1}{\sqrt{2^{n_{s}}\sqrt{\pi}l_{c}}}H_{n_{s}}\left(\frac{z}{l_{c}}\right)e^{-\frac{z^{2}}{2l_{c}^{2}}}
\end{array}\label{eq:A18}
\end{equation}

which is essentially the electron bunch distribution modulated with
the Hermite Gauss Super Mode basis given in eq. (\ref{eq:A6}). In
Fig.14 we have reported the comb structure $S(z)$ (eq. (\ref{eq:A18})
) and it results that the agreement between eq. (\ref{eq:A18}) and
the numerical results is more than reasonable.

\begin{figure}
\includegraphics{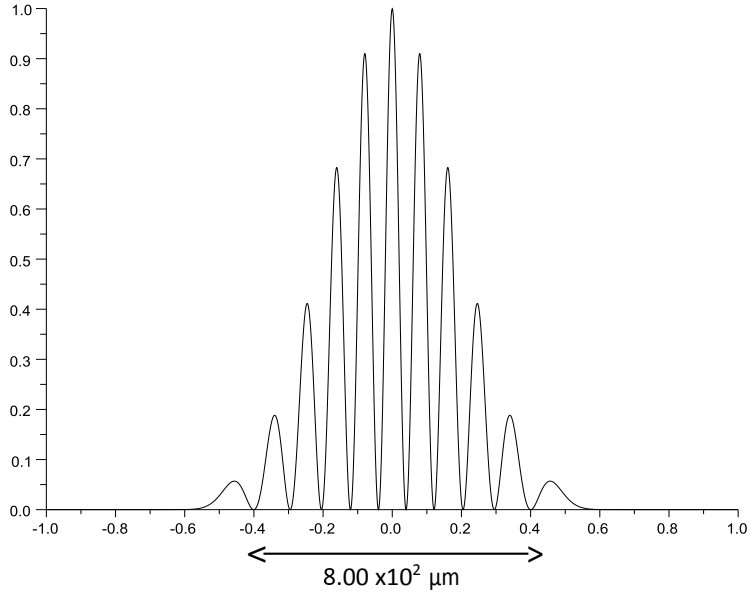}

\protect\caption{Comb structure (eq. (\ref{eq:A18}) ) to be compared with Fig.\ref{Fig.3}}

\end{figure}

\end{document}